 \documentclass[twocolumn,aps,prb,floatfix]{revtex4}
\usepackage{graphicx,graphics,color,epsfig}
\usepackage{amsmath}
\usepackage{amssymb}
\usepackage[small]{caption}

\begin{document}
\title{
Nonequilibrium dynamic exponent and spin-glass transitions
}
\author{Tota Nakamura}
\affiliation{%
Laboratory of Natural Science,
Shibaura Institute of Technology, %
307 Fukasaku, Minuma, Saitama 337-8570, Japan}

\date{ }

\begin{abstract}
Nonequilibrium dynamics of 
the $\pm J$ Ising, the {\it XY}, and the Heisenberg spin-glass
models are investigated in three dimensions.
A nonequilibrium dynamic exponent is calculated from the
dynamic correlation length.
The spin-glass dynamic exponent continuously depends on the temperature.
There is no anomaly at the critical temperature as is recently reported by
Katzgraber and Campbell.
On the other hand, the chiral-glass dynamic exponent takes a constant value 
above the spin-glass transition temperature ($T_\mathrm{sg}$),
and becomes temperature-dependent below $T_\mathrm{sg}$.
The finite-time scaling analyses on the spin- and the chiral-glass 
susceptibility 
are performed using the temperature dependence of the dynamic exponent.
The critical temperature and the critical exponents are estimated with
high precisions.
In the Ising model,
$T_\mathrm{sg}=1.18(2)$, $\nu_\mathrm{sg}=1.75(10)$, 
$\gamma_\mathrm{sg}=3.4(2)$, and 
$z_\mathrm{sg}=5.1(1)$.
In the {\it XY} model,
$T_\mathrm{sg}=0.435(15)$, $\nu_\mathrm{sg}=1.25(15)$, 
$\gamma_\mathrm{sg}=2.30(25)$, and $z_\mathrm{sg}=5.0(1)$ 
for the spin-glass transition,
and
$T_\mathrm{cg}=0.445(15)$, $\nu_\mathrm{cg}=1.05(10)$, 
$\gamma_\mathrm{cg}=1.65(20)$, and $z_\mathrm{cg}=5.1(2)$ 
for the chiral-glass transition.
In the Heisenberg model,
$T_\mathrm{sg}=0.18(1)$, $\nu_\mathrm{sg}=1.55(15)$, 
$\gamma_\mathrm{sg}=2.9(2)$, and $z_\mathrm{sg}=5.1(1)$ 
for the spin-glass transition,
and
$T_\mathrm{cg}=0.23(1)$, $\nu_\mathrm{cg}=0.70(15)$, 
$\gamma_\mathrm{cg}=0.65(15)$, and $z_\mathrm{cg}=5.0(2)$ 
for the chiral-glass transition.
A difference of the spin- and the chiral-glass transition temperatures 
is resolved in the Heisenberg model.
The dynamic critical exponent takes almost the same value for all transitions.
It suggests that the spin-glass and the chiral-glass transitions 
in three dimensions are dynamically universal.

\end{abstract}
\maketitle

\section{Introduction}
\label{sec:intro}

Randomness and frustration produce nontrivial phenomena in
spin glasses\cite{SGReview1,SGReview2,SGReview3}.
There is no spatial order in the spin-glass phase.
The spins are ordered only in the time direction:
spins randomly freeze below the critical temperature.
Growth of the spin-glass correlations is very slow.
Most of interesting
observations in experiments and in computer simulations are 
essentially nonequilibrium phenomena.
Aging, memory and rejuvenation effects are the typical examples.%
\cite{dupuis1,jonsson1,bouchaud1,bert1}
It has become clear that a domain size of frozen spins plays an important role
in explaining these phenomena.%
\cite{fisherhuse,bouchaud2,yoshino1,berthier1,scheffler,yoshino2,parisi1,%
kisker1,komori1,berthier2,dupuis2,jonsson2,yoshino3,nakamura3,%
berthier3,katzgraber}
Details of time evolution of the domain size may determine the dynamics of
spin glasses.

In the second-order phase transitions,
a correlation length $\xi$ and a correlation time $\tau$ 
diverge at the critical temperature.
The dynamic scaling relation holds as $\tau \sim \xi^{z}$
with the dynamic critical exponent $z$.
If we observe the correlation length as a function of time,
it also diverges algebraically with the same exponent.
It is written as
$
\xi(t) \sim t^{1/z}.
$
Consequently, the magnetic susceptibility $\chi(t)$ 
diverges with an exponent $\gamma/z\nu$.

The dynamic scaling relation also holds in spin glasses at the critical 
temperature.
Huse\cite{huse}
remarked that a time evolution of the spin-glass susceptibility 
$\chi_\mathrm{sg}(t)$ diverges 
algebraically with an exponent $\gamma/z\nu$ in the Ising spin-glass model.
Blundell {\it et al.}\cite{blundell} found that
a relaxation function of the Binder parameter $g_\mathrm{sg}$ also diverges
with an exponent $1/z$.
The spin-glass correlation length is also considered to diverge algebraically
at the critical temperature.

Below the spin-glass transition temperature, $T_\mathrm{sg}$, 
the algebraic divergence of the spin-glass correlation length is considered to 
continue in the nonequilibrium process.
An exponent of the divergence linearly depends
on the temperature:\cite{berthier2,parisi1,kisker1,komori1,yoshino3,nakamura3}
\begin{equation}
\frac{1}{z_\mathrm{s}(T)} = \frac{T}{z_\mathrm{sg} T_\mathrm{sg}}.
\label{eq:zne}
\end  {equation}
Here, we call $z_\mathrm{s}(T)$ a spin-glass nonequilibrium dynamic exponent.
It is defined only in the nonequilibrium relaxation process.
It equals to the dynamic critical exponent of the spin glass transition,
$z_\mathrm{sg}$, at the critical temperature.

Recently, Katzgraber and Campbell \cite{katzgraber}
remarked that the relation (\ref{eq:zne}) 
continues to hold above $T_\mathrm{sg}$, and that
there is no anomaly at the critical temperature.
The model they investigated is the Gaussian-bond
Ising spin-glass model and the gauge-glass model in three and four dimensions.
It is very important whether or not this finding is common to various
spin-glass models.
In regular systems, the dynamic critical exponent $z$ 
governs the algebraic growth of the correlation length in the nonequilibrium
relaxation process above the critical temperature.
The temperature dependence of $z$ above $T_\mathrm{sg}$ is out of our
assumption regarding the dynamics of spin glasses.

The dynamic scaling relation in the nonequilibrium relaxation process of the 
spin-glass models suggests that we can study the spin-glass phase transition 
using the nonequilibrium relaxation method.\cite{Stauffer,Ito,ItoOz1,OzIto2}
It is a powerful tool to estimate the critical temperature and the critical
exponent with high precisions.
It has been applied to slow dynamic systems with
frustration and/or randomness.\cite{OzIto2,shirahata1,nakamura,nakamura2,%
shirahata2,yamamoto1,nakamura3,nakamura4}
A key concept of the method is an exchange of time and size by the
dynamic scaling relation, $t \rightleftarrows L^z$.
Here, a single exponent $z$ is considered to satisfy the relation above the
critical temperature.
This assumption may be broken by the relation (\ref{eq:zne}).
Therefore, 
we must reconsider the nonequilibrium relaxation analysis in the spin-glass 
problems taking into account the relation (\ref{eq:zne}).
Then, the finite-time-scaling results may be influenced.

In this paper we carry out extensive Monte Carlo simulations. 
It is made clear
that the relation (\ref{eq:zne}) is satisfied for the spin-glass transitions
in all models.
It continues regardless the critical temperature without anomaly.
On the other hand, 
the temperature dependence of the nonequilibrium dynamic exponent of the
chiral-glass correlation length only appears below $T_\mathrm{sg}$.
It takes a constant value above $T_\mathrm{sg}$ as in the regular systems.
The finite-time scaling analyses are performed using the temperature dependence
of the dynamic exponent.
The transition temperature and the critical exponents are estimated
with higher precisions compared to our previous results.
We give a conclusion to a problem of whether or not 
the spin- and the chiral-glass transitions occur simultaneously
in the {\it XY} and the Heisenberg models.

The present paper is organized as follows.
Section \ref{sec:model} describes a model
and physical quantities we observe in simulations.
Section \ref{sec:method} explains our numerical method and some technical
details.
Results of the Ising, the {\it XY}, and the Heisenberg models are presented in 
Sec. \ref{sec:isingresults}, \ref{sec:xyresults}, and \ref{sec:Hsgresults},
respectively.
Section \ref{sec:discussion} is devoted to summary and discussions.

\section{Model and Observables}
\label{sec:model}

\subsection{Model}

We study the $\pm J$ spin-glass model in a simple cubic lattice.
\begin{equation}
  {\cal H} = - \sum_{\langle ij \rangle} 
     J_{ij} \mbox{\boldmath $S$}_i \cdot 
            \mbox{\boldmath $S$}_j 
\label{eq:H1}
\end{equation}
The sum runs over all the nearest-neighbor spin pairs $\langle ij \rangle$.
The interactions, $J_{ij}$, take the two values  $+J$ and $-J$
with the same probability.
The temperature, $T$, is scaled by $J$.
Linear lattice size is denoted by $L$. 
A total number of spins is $N = L \times L \times (L+1)$, and
skewed periodic boundary conditions are imposed.
The spins are updated by the single-spin-flip algorithm
using the conventional Metropolis probability.
We treat three types of spin dimensions: the Ising model with
$\mbox{\boldmath $S$}_i=S_{iz}$, 
the {\it XY} model with $\mbox{\boldmath $S$}_i=(S_{ix}, S_{iy})$,
and the Heisenberg model with 
$\mbox{\boldmath $S$}_i=(S_{ix}, S_{iy}, S_{iz})$.
In the {\it XY} model, each spin is defined by an angle in the {\it XY} plane:
$\mbox{\boldmath $S$}_i = (\cos\theta_i,\sin\theta_i)$.
The angle $\theta _i$ takes continuous values in $[0,2\pi)$; however,
we divide the interval into 1024 discrete states in this study.
The effect of the discretization is checked by comparing data
of 1024-state simulations with those of 2048-state simulations.
It is found to be negligible.

The physical quantities that are calculated in our simulations are 
the spin-glass susceptibility, $\chi_\mathrm{sg}$,
the chiral-glass susceptibility, $\chi_\mathrm{cg}$, and
the spin- and the chiral-glass correlation functions,
$f_\mathrm{sg}$ and $f_\mathrm{cg}$,
from which we estimate the correlation lengths, 
$\xi_\mathrm{sg}$ and $\xi_\mathrm{cg}$.

\subsection{Spin-glass susceptibility}

The spin-glass susceptibility is defined by the following expression.
\begin{equation}
\chi_\mathrm{sg}\equiv\frac{1}{N}
\left[
\sum_{i,j}\langle \mbox{\boldmath $S$}_i \cdot \mbox{\boldmath $S$}_j \rangle^2
\right]_\mathrm{c}
\label{equ:xsg}
\end{equation}
The thermal average is denoted by $\langle \cdots \rangle$, and 
the random-bond configurational average is denoted by $[ \cdots ]_\mathrm{c}$.
The thermal average is replaced by an average over independent real
replicas that consist of different thermal ensembles:  
\begin{equation}
\langle \mbox{\boldmath $S$}_i \cdot \mbox{\boldmath $S$}_j \rangle
= \frac{1}{m}\sum_{A=1}^m 
\mbox{\boldmath $S$}_i^{(A)} \cdot \mbox{\boldmath $S$}_j^{(A)}.
\end  {equation}
A replica number is denoted by $m$.
The superscripts $A$ is the replica index.
We prepare $m$ real replicas for each random bond 
with different initial spin configurations.
Each replica is updated using a different random number sequence.
The thermal average is taken only by this replica average in our numerical
scheme.
A replica number controls an accuracy of the thermal average.

An overlap between two replicas, $A$ and $B$, is defined by
\begin{equation}
  q_{\mu \nu}^{A B} \equiv \frac{1}{N}
   \sum _i {S}_{i\mu}^{(A)} {S}_{i\nu}^{(B)}.
\label{equ:qsg}
\end{equation}
Here, subscripts $\mu$ and $\nu$ represent three spin directions 
$x, y$, and $z$.
The spin-glass susceptibility is estimated using the overlap function:
\begin{equation}
\chi_\mathrm{sg} = \frac{2N}{m(m-1)}
\left[
\sum_{A>B, \mu,\nu}
\langle
(q^{AB}_{\mu\nu})^2
\rangle
\right]_\mathrm{c}.
\end  {equation}
Similarly, the chiral-glass susceptibility in the {\it XY} model
\cite{KawamuraXY1,KawamuraXY2,KawamuraXY3} is defined by 
\begin{equation}
  \chi_\mathrm{cg} \equiv \frac{6N}{m(m-1)}
\left[
\sum_{A > B}
(q_{\kappa}^{A B})^2
\right]_\mathrm{c} ,
\label{equ:xcg}
\end{equation}
where
\begin{eqnarray}
  q_{\kappa}^{A B}&\equiv&\frac{1}{3N}\sum _{\alpha} \,
     {\kappa}_{\alpha}^{(A)} {\kappa}_{\alpha}^{(B)},
\label{equ:qcg}
\\
 \kappa _{\alpha}^{(A)}&\equiv&\frac{1}{2 \sqrt{2}} 
 \sum_{\langle jk\rangle\in\alpha} J_{jk} 
 \sin{( \theta _j^{(A)} - \theta _k^{(A)} )}.
\label{equ:localk}
\end{eqnarray}
A local chirality variable, $\kappa_{\alpha}^{(A)}$, is 
the $z$ component of the vector chirality.
It is defined at each square plaquette, $\alpha$, 
that consists of the nearest-neighbor bonds, $\langle jk\rangle$. 
In the Heisenberg model, we use the scalar 
chirality\cite{KawamuraH1,HukushimaH,HukushimaH2}, which is defined by
\begin{equation}
 \kappa _{i,\mu}^{(A)}\equiv
\mbox{\boldmath $S$}_{i+\hat{\mbox{\boldmath $e$}}_{\mu}}^{(A)}
\cdot
(
\mbox{\boldmath $S$}_{i}^{(A)}
\times
\mbox{\boldmath $S$}_{i-\hat{\mbox{\boldmath $e$}}_{\mu}}^{(A)}),
\end  {equation}
where $\hat{\mbox{\boldmath $e$}}_{\mu}$ denotes a unit lattice vector along
the $\mu$ axis.
The overlap function and the chiral-glass susceptibility 
are defined in the same manner as in the {\it XY} model.

\subsection{Dynamic correlation length}

A spin-glass correlation function is defined by the following expression
(\ref{eq:sgcordef}).
\begin{eqnarray}
f_\mathrm{sg}(r) &\equiv& \left[\sum_i^N\langle
 \mbox{\boldmath $S$}_i \cdot \mbox{\boldmath $S$}_{i+r} \rangle^2\right]_\mathrm{c}
\label{eq:sgcordef}
\\
&=& \left[ 
\frac{2}{m(m-1)}
\sum_{A>B, i, \mu\nu}
q_{\mu\nu}^{AB}(i)
q_{\mu\nu}^{AB}(i+r)
\right]_\mathrm{c}
\label{eq:sgcorover}
\\
&=&\left[
\sum_i^N
\left(
\frac{1}{m} 
\sum_A^m
\mbox{\boldmath $S$}_i^{(A)} \cdot \mbox{\boldmath $S$}_{i+r}^{(A)}
\right)^2
\right]_\mathrm{c}
\label{eq:sgcoruse}
\end{eqnarray}
Here,
$
q_{\mu\nu}^{AB}(i)=
S_{i\mu}^{(A)}S_{i\nu}^{(B)}
$
is the spin overlap on the $i$th site.
When a replica number is two, it is equivalent to 
the four-point correlation function as is shown in Eq.~(\ref{eq:sgcorover}).
Since $m$ is large in this study, it is very time-consuming to take an 
average over $m(m-1)/2$ different overlap functions.
Therefore, we use another expression (\ref{eq:sgcoruse}) for $f_\mathrm{sg}(r)$.
For a given distance $r$ and a site $i$, we calculate a spin correlation 
function for replica $A$, and store it in a memory array.
Then, the replica average is taken and the obtained value is squared.
Taking a summation of the squared value over lattice sites $i$ 
we obtain the correlation function $f_\mathrm{sg}(r)$ for a given distance $r$.
Changing a value of $r$ we do the same procedure to evaluate $f_\mathrm{sg}(r)$.
Total amount of calculation is much smaller in this procedure because
the maximum value of $r=L/2$ is much smaller than $m(m-1)/2$.

We obtain a chiral-glass correlation function in the same manner
replacing the local spin variables with the local chirality variables.
Here, we only consider correlations between two parallel square plaquettes
in the {\it XY} model.
A pair of three neighboring spins in the same direction is considered 
in the Heisenberg model.
We estimate the spin- and the chiral-glass correlation lengths, 
$\xi_\mathrm{sg}$ and $\xi_\mathrm{cg}$, from the correlation functions
using the following expression.
\begin{equation}
f_\mathrm{sg/cg}(r)  = f_0
\exp\left( -\frac{r}{\xi_\mathrm{sg/cg}}\right)
\label{eq:fsg}
\end{equation}

A problem in this expression is a prefactor $f_0$.
We may suppose several forms for $f_0$.
For example, $f_0\sim l^{2-d-\eta}$.
However, the obtained value may depend on a value of $\eta$, 
which is not known yet.
Therefore, we take a rather brute-force approach as explained below.
Then, the relaxation functions of the correlation length,
$\xi_\mathrm{sg}$ and $\xi_\mathrm{cg}$, are obtained.
This is sometimes called the dynamic correlation length.

\begin{figure}
  \begin{center}
  \resizebox{0.45\textwidth}{!}{\includegraphics{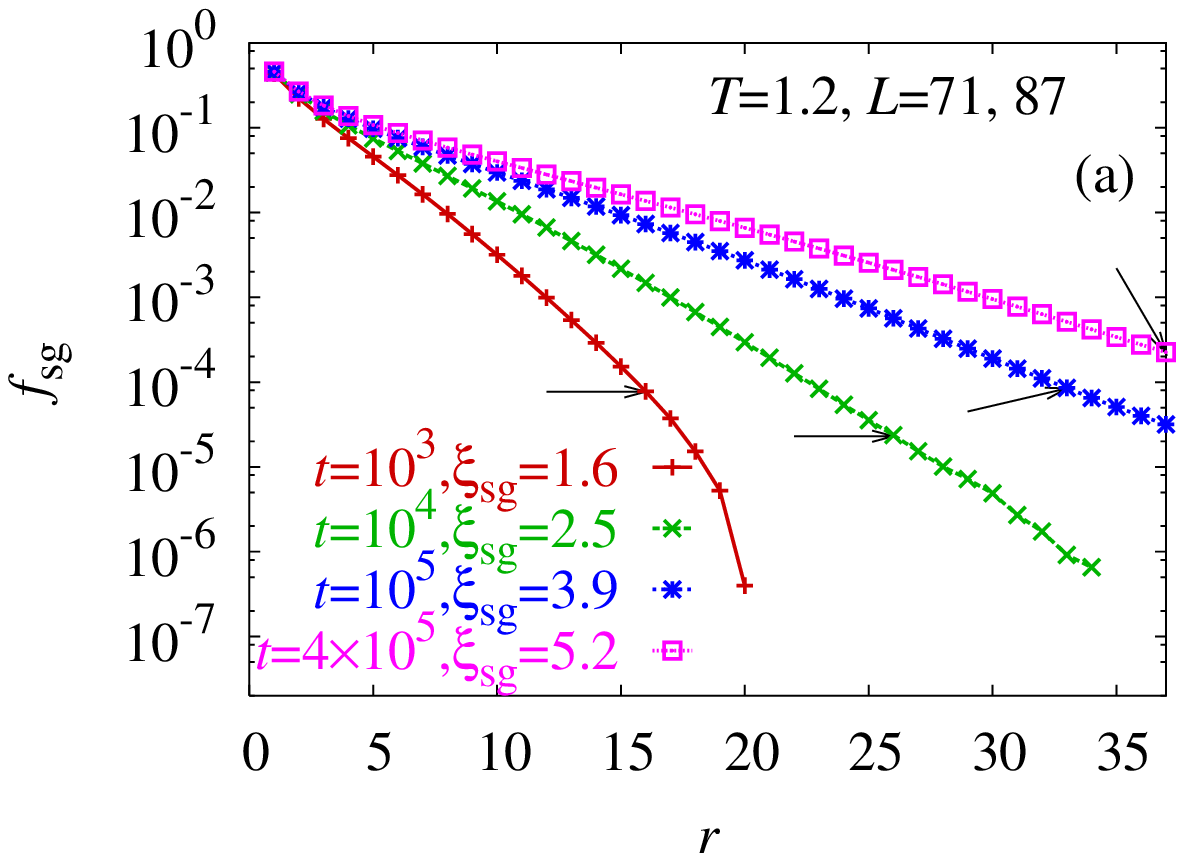}}
  \resizebox{0.45\textwidth}{!}{\includegraphics{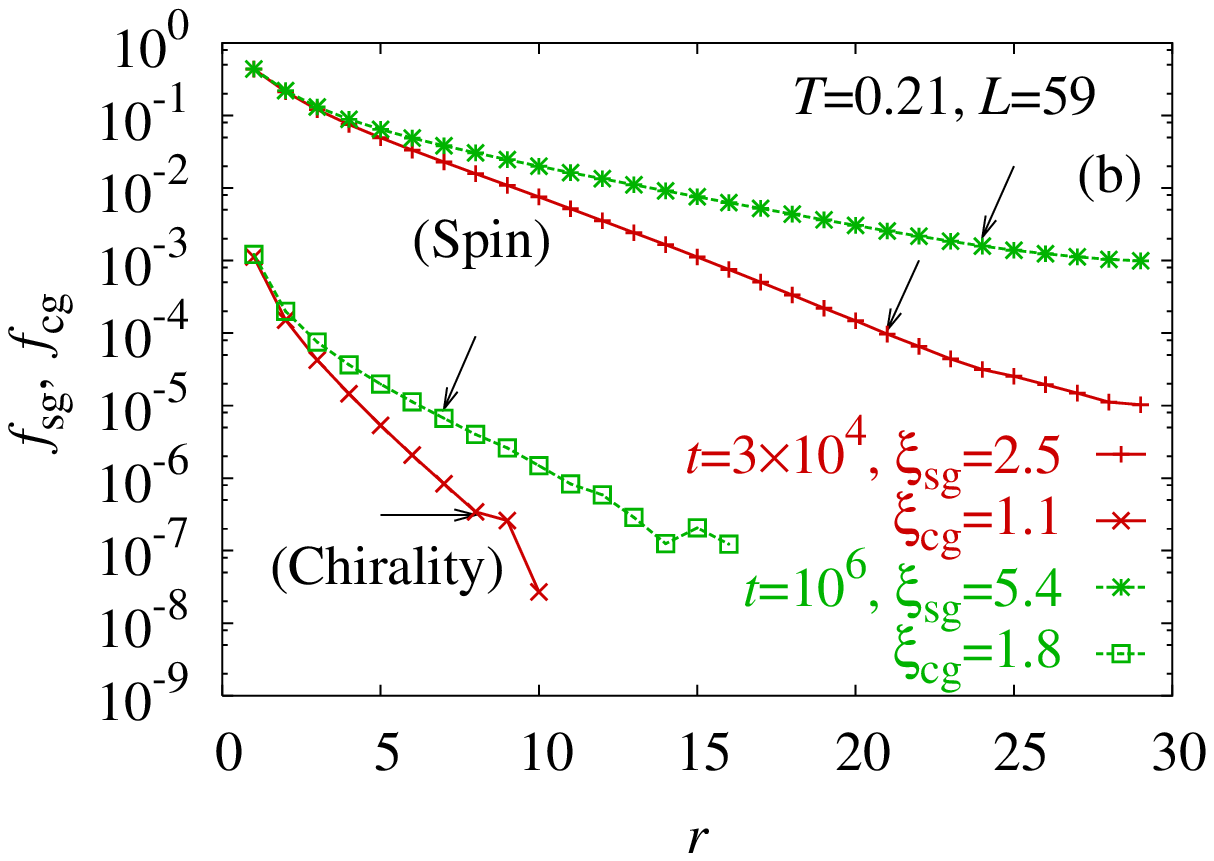}}
  \end{center}
  \caption{
(Color online)
Spin- and chiral-glass correlation functions are plotted versus distance $r$ 
at various time step $t$.
(a) Results of the Ising model.
The temperature $T=1.2$ is near the critical point. 
The lattice size $L=71$ for $t=10^3$ and
$L=87$ for $t\ge 10^4$.
(b) Results of the Heisenberg model.
The temperature $T=0.21$ is near the chiral-glass transition temperature.
The lattice size $L=59$.
Arrows depict a cut-off distance when estimating the correlation length. 
}
\label{fig:fsgising}
\end{figure}

We prepare a sufficiently large lattice and calculate the correlation 
functions with high precisions, typically up to an order of $10^{-6}$.
The correlation functions asymptotically exhibit a single exponential decay 
when $r$ is large enough.
We estimate the correlation length using data in this asymptotic region.
Figure \ref{fig:fsgising} shows the lin-log plot of $f_\mathrm{sg}(r)$ 
and $f_\mathrm{cg}(r)$ in the Ising and the Heisenberg models 
near the critical temperature.
Linearity is good when the distance is larger than a typical distance,
$r_\mathrm{min}$. 
This behavior guarantees our single-exponential fitting when
estimating the correlation lengths.
Values of $r_\mathrm{min}$ are summarized in Sec. \ref{sec:sim-cond}.

Correlation functions sometimes exhibit bending to convergence as shown
in Fig.~\ref{fig:fsgising}(b).
It is considered to be an outcome of the boundary effect.
Since we work with the skewed periodic boundary conditions, the farthest 
distance is $L/2$.
The correlation data near $r=L/2=29$ are 
influenced by the boundary conditions.
Therefore, the spin-glass data when $r>25$ show convergence.
We cut off such data when we estimate the correlation length.
We also cut off data which fluctuate from the fitting line.
The minimum distance and the cut-off distance are determined 
so that the relative deviations of data
from the fitting line stay within 5\%.
Arrows in Fig.~\ref{fig:fsgising} depict the cut-off distance.

It is also noted that the chiral-glass correlation functions are smaller than
the spin-glass ones by three digits.
Estimations of the chiral-glass correlation length are usually very difficult,
and the error bars become relatively large.

\section{Method}
\label{sec:method}

\subsection{Nonequilibrium relaxation method}

We start the simulations with random spin configurations, which 
is a paramagnetic state at $T=\infty$.
The temperature is quenched to a finite value at the first Monte Carlo step.
We calculate the physical observables at each observation time, 
and obtain the relaxation functions.
Changing the initial spin state, the random bond configuration, 
and the random number sequence, we start
another simulation to obtain another set of the relaxation functions.
We calculate the average of the relaxation functions
over these independent Monte Carlo simulations.

It is necessary to check a time scale 
when the finite-size effects appear in the relaxation functions.
We must stop the simulations before the size effects appear.
The nonequilibrium relaxation method is based upon 
taking the infinite-size limit before the equilibrium limit is taken.
Data presented in this paper can be regarded as those of the infinite-size
system.

\subsection{Finite-time scaling analysis}
\label{sec:scalingprocedure}

We estimate the transition temperature using the
finite-time-scaling analysis.\cite{OzIto2}
This is a direct interpretation of the conventional 
finite-size-scaling analysis considering the dynamic scaling 
hypothesis: $\tau \sim \xi^{z}$.
Through this relation, the ``size'' is replaced by the ``time.''
Here, the temperature dependence of the nonequilibrium dynamic exponent,
$z_\mathrm{s}(T)$, written in Eq.~(\ref{eq:zne}) should be taken into account.
The temperature coefficient, $1/z_\mathrm{sg}T_\mathrm{sg}$, 
is obtained beforehand by the scaling analysis of the
dynamic correlation length.
The following equation represents the scaling formula for the spin-glass
susceptibility.
\begin{equation}
\chi_\mathrm{sg}(t) t^{-\gamma_\mathrm{sg}/z_\mathrm{s}(T) \nu_\mathrm{sg}} =  
  \tilde{\chi}_\mathrm{sg}(t/
(T-T_\mathrm{sg})^{-z_\mathrm{s}(T)\nu_\mathrm{sg}})
\label{equ:xsg_scl}
\end{equation}
Here, $\tilde{\chi}$ denotes a universal scaling function.
We plot data of the spin-glass susceptibility multiplied by 
$t^{-\gamma_\mathrm{sg}/z_\mathrm{s}(T)\nu_\mathrm{sg}}$ against 
$t/(T-T_\mathrm{sg})^{-z_\mathrm{s}(T)\nu_\mathrm{sg}}$. 
We choose scaling parameters, $T_\mathrm{sg}$, $\gamma_\mathrm{sg}$, 
and $\nu_\mathrm{sg}$ so that
they yield the best universal scaling function.

We perform a scaling analysis using a criterion 
which we proposed in the previous paper.\cite{yamamoto1}
In general,
it is difficult to choose proper scaling parameters because the 
universal scaling function form is not known.
One usually approximate the scaled data by some polynomial functions,
and choose scaling parameters so that the fitting error becomes smallest.
However, if we change the scaling parameter, the function form itself changes.
Then, a significance of the fitting error changes.
A value of the fitting error sometimes misleads us to incorrect results.
Therefore, we take another strategy.

Since the correct scaling function is universal, it is free from human factors.
It should be robust against a change of a temperature range 
of data we use in the analysis;
it should be robust against a change of discarding steps in simulations;
it should be robust against a change of procedures determining scaling
parameters.
In this paper, we consider the following three procedures:
\begin{itemize}
\item (Procedure-1)
For a given value of $\gamma_\mathrm{sg}/\nu_\mathrm{sg}$, 
search $T_\mathrm{sg}$ and
$\nu_\mathrm{sg}$ that yield the smallest fitting error.
\item (Procedure-2)
For a given value of $\gamma_\mathrm{sg}$, search $T_\mathrm{sg}$ and
$\nu_\mathrm{sg}$ that yield the smallest fitting error.
\item (Procedure-3)
For a given value of $\nu_\mathrm{sg}$, search $T_\mathrm{sg}$ and
$\gamma_\mathrm{sg}$ that yield the smallest fitting error.
\end  {itemize}
Here, we mean the smallest fitting error by the least-square deviations when
we approximate the scaled data with a polynomial function of seven orders.
We also change the temperature range and a number of discarding steps
for each procedure.
Scaling results systematically depend on these procedures.
The most probable result is obtained by a crossing intersection or a merging
point because it should be independent of the procedures.
An error bar of the result is given by a range of scaling parameters 
which yield a good scaling plot.

\subsection{Replica number VS Sample number}

A replica number controls an accuracy of the thermal average.
A sample number controls an accuracy of the configurational average.
In order to improve an accuracy of the thermal average, $m$ should be large.
However, if we set it large, a sample number is restricted, which causes
a poor configurational average.
So, there is a question.
In order to improve an accuracy of the final results,
which number should be set large first, a replica number $m$ or
a sample number $N_\mathrm{s}$?
Most of the simulational studies in spin glasses take a choice of $m=2$.

\begin{figure}
  \begin{center}
  \resizebox{0.45\textwidth}{!}{\includegraphics{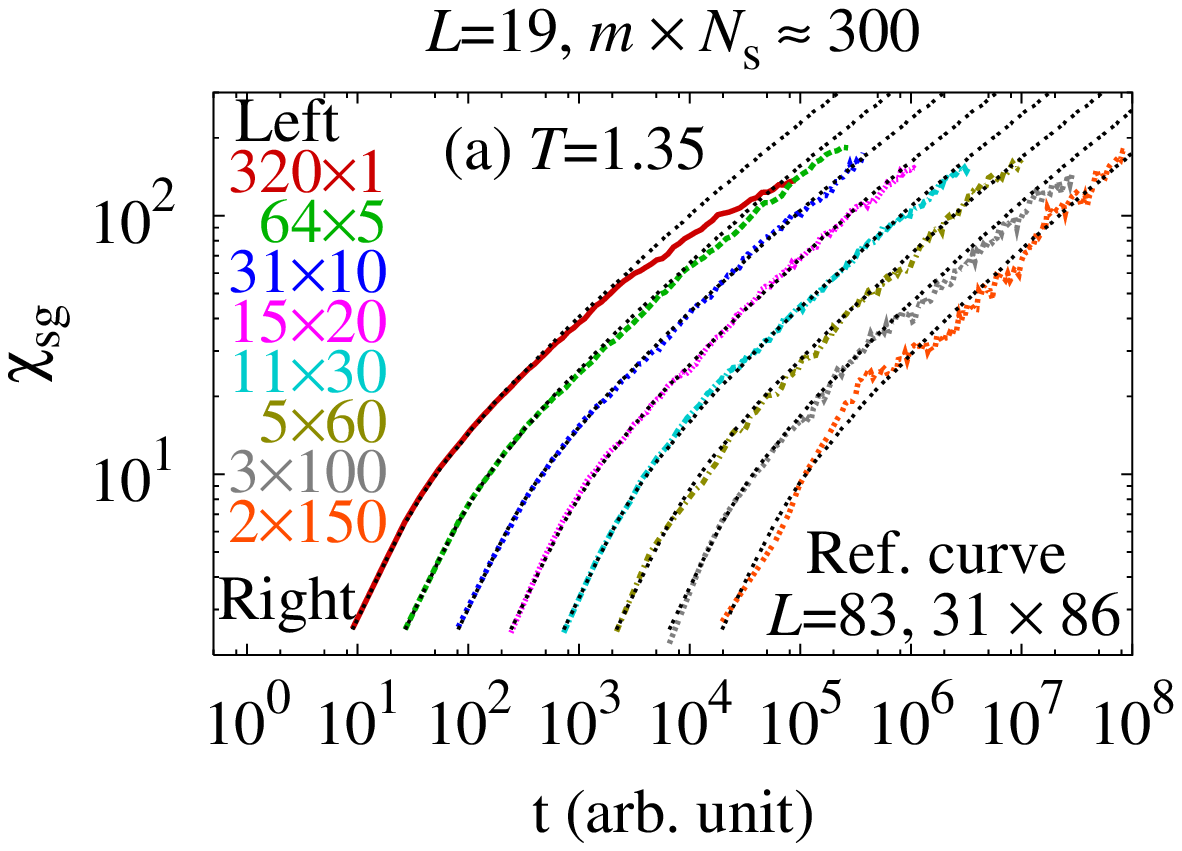}}
  \resizebox{0.45\textwidth}{!}{\includegraphics{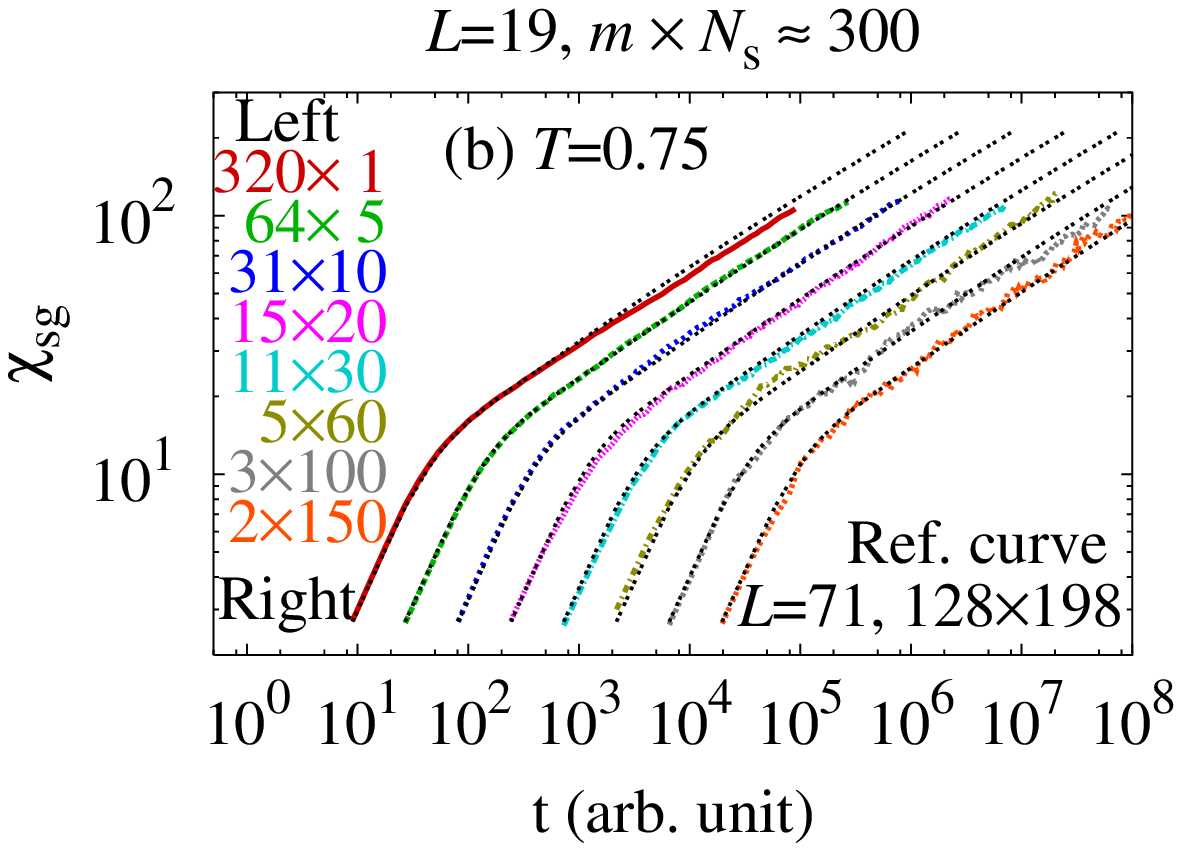}}
  \end{center}
  \caption{
(Color online)
Relaxation functions of the spin-glass susceptibility of the Ising model
when a replica number $m$ and a bond sample number $N_\mathrm{s}$
are changed keeping the product almost constant, $m\times N_\mathrm{s}\sim 300$.
(a) $T=1.35 >T_\mathrm{c}$, and (b) $T=0.75 <T_\mathrm{c}$.
Reference curves are plotted with black dotted line, which are obtained
by (a) $L=83$, $m=31$, and $N_\mathrm{s}=86$;
(b) $L=71$, $m=128$, and $N_\mathrm{s}=198$.
Each function is shifted to the $t$ direction in order to compare the
deviations.
}
\label{fig:nrepconf}
\end{figure}
Figure \ref{fig:nrepconf} shows the answer.
Relaxation functions of the spin-glass susceptibility in the Ising model 
are plotted both in the paramagnetic phase(a) and in the spin-glass phase(b).
A product $m \times N_\mathrm{s}$ is set almost equal to 300 and
the lattice size is set to $L=19$.
A total number of spin updates and a total computational time are same.
We also plot a reference curve which is obtained by a sufficiently large 
simulation.
All relaxation curves of $L=19$ are consistent with the reference curves.
Statistical fluctuations are large when the replica number is small.
They are suppressed as we increase a replica number and
decrease a sample number.
However, when a sample number is one, there is a systematic deviation from
the reference curve.
This deviation depends on each random-bond sample.
The dependence becomes small and the deviation appears later if we increase
the lattice size.
A sample number can be set small in the nonequilibrium process of a 
sufficiently large system, where the size can be considered as infinity.
Therefore,
a replica number should be set larger first when the total
computational time is restricted.
We need at least several sample numbers to avoid systematic deviation due to
the sample dependence.
The statement is true both in the paramagnetic phase 
and in the spin-glass phase.

In this study, we set $m=128$ and $m=192$ for the Ising model,
$m=32$ and $m=64$ for the {\it XY} model,
and $m=32$ for the Heisenberg model.
Simulation conditions are summarized in Sec. \ref{sec:sim-cond}.

\subsection{Updating probability: Metropolis VS Heat-bath}

\begin{figure}
  \begin{center}
  \resizebox{0.45\textwidth}{!}{\includegraphics{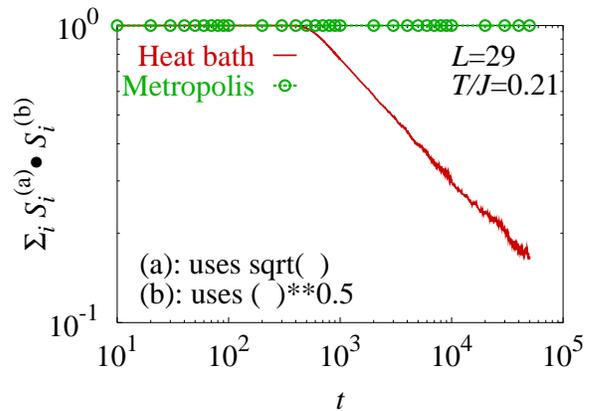}}
  \end{center}
  \caption{
(Color online)
A replica overlap in the Heisenberg model
between the exactly same replica with the same random 
number sequence but with a different usage of a square-root function.
Replica (a) uses a built-in sqrt function, and
replica (b) uses an exponential function. 
A heat-bath algorithm and a Metropolis algorithm are compared.
A temperature, $T/J=0.21$, is near the chiral-glass transition point.
}
\label{fig:chaos}
\end{figure}

Spin update dynamics in this paper is the Metropolis algorithm.
A heat-bath algorithm \cite{Olive}
is usually adopted in a simulation of the Heisenberg model.
However, we found a chaotic behavior in a nonequilibrium dynamics 
of the heat-bath algorithm as shown in Fig. \ref{fig:chaos}. 

We prepare two real replicas, (a) and (b),  
with the same bond configurations with
the same initial spin configurations with the same random number
sequence.
Only one difference is a use of a square-root function.
A replica-(a) uses a built-in {\tt sqrt(x)} function for $\sqrt{x}$, while
a replica-(b) uses an exponential function {\tt x**0.5}.
An spin overlap function between two replicas is observed.
Mathematically, it must stay unity forever.
However, the overlap decays after some characteristic time
in the heat-bath algorithm.
This time depends on the temperature and also a compiler option which
controls numerical precisions.
It becomes longer for higher precisions.
To the contrary,
an overlap in the Metropolis algorithm stays unity.

Mathematical functions are numerically approximated in computer calculations.
A square-root function, $\sqrt{x}$, 
is one of the most difficult function to approximate.
The error is negligible for each function call.
However, successive calls may enlarge the error. 
In the heat-bath algorithm,\cite{Olive} a spin state, 
$\mbox{\boldmath $S$}_i$, is a function of
a local molecular field from the neighboring spins, 
$\mbox{\boldmath $S$}_j$.
The neighboring spin state, $\mbox{\boldmath $S$}_j$, is also 
a function of the molecular field which contains $\mbox{\boldmath $S$}_i$.
Therefore, a spin state $\mbox{\boldmath $S$}_i$ given by a complex function of
spin states of itself and neighboring spins in the past.
A square-root function is called successively in this procedure.
Therefore, a small numerical error may be enhanced.
A spin state may become different from the correct one.
We cannot know which state is correct because every computer simulation
contains a numerical error.

In the Metropolis algorithm, each spin state is generated independently
by random numbers.
There is no correlation between spin-state candidates in a time sequence.
The overlap in Fig. \ref{fig:chaos} stays unity as mathematically required.
In this paper, we study a spin-chirality separation 
in the Heisenberg model.
The separation may be observed in a very narrow temperature range.
We deal with this very naive problem by the relaxation method.
Therefore,
we use a Metropolis algorithm
in order to exclude out the unphysical chaotic behavior 
from the relaxation process.

\subsection{Simulation conditions}
\label{sec:sim-cond}

In this subsection, we summarize simulation conditions in 
Tabs.
\ref{tab:isg1}, \ref{tab:isg2}, \ref{tab:xy1}, \ref{tab:xy2}, and \ref{tab:hsg}.
A sample number $N_\mathrm{s}$, a linear size $L$, a minimum distance
of the spin correlations $r_\mathrm{min}$ vary with a Monte Carlo time step.
Numbers above right arrows in tables 
denote a time step when the condition changes.

In the Ising model,
a replica number is 128 except for a case of $T=1.4$ and $L=111$, 
where it is 192.
Simulation conditions when the dynamic correlation length is estimated are
listed in Tab. \ref{tab:isg1}, and
those when the spin-glass susceptibility is calculated are listed in
Tab. \ref{tab:isg2}.

\begin{table}
\begin{tabular}{ c c c c }
\hline
\hline
$T$ & $L$ & $r_\mathrm{min}$ & $N_\mathrm{s}$ \\
\hline
1.7 & 71 &  7 & 98 \\
1.6 & 71 $\longrightarrow^{\hspace*{-0.5cm} 2\cdot 10^3}$ 87
    &  7 $\longrightarrow^{\hspace*{-0.5cm} 2\cdot 10^3}$ 12
    & 659 $\longrightarrow^{\hspace*{-0.5cm}5\cdot 10^3}$ 651 \\
1.5 & 71 $\longrightarrow^{\hspace*{-0.5cm}10^4}$ 95 
    & 10 $\longrightarrow^{\hspace*{-0.5cm}10^4}$ 16
    & 309 $\longrightarrow^{\hspace*{-0.5cm}500}$ 95 
          $\longrightarrow^{\hspace*{-0.5cm}10^4}$ 338 \\
1.4 & 71 $\longrightarrow^{\hspace*{-0.5cm} 5\cdot 10^4}$ 111
    & 10 $\longrightarrow^{\hspace*{-0.5cm} 5\cdot 10^4}$ 16
    & 426 $\longrightarrow^{\hspace*{-0.5cm} 5000}$  232
          $\longrightarrow^{\hspace*{-0.5cm} 5\cdot 10^4}$ 25 \\
1.3 & 71  $\longrightarrow^{\hspace*{-0.5cm} 10^4}$ 87
    & 10  $\longrightarrow^{\hspace*{-0.5cm} 10^4}$ 16
    & 112 $\longrightarrow^{\hspace*{-0.5cm} 10^4}$ 48 \\
1.2 & 71  $\longrightarrow^{\hspace*{-0.5cm} 10^4}$ 87
    & 10  $\longrightarrow^{\hspace*{-0.5cm} 10^4}$ 14
    & 236 $\longrightarrow^{\hspace*{-0.5cm} 10^4}$ 163
          $\longrightarrow^{\hspace*{-0.5cm} 5\cdot 10^4}$ 85
          $\longrightarrow^{\hspace*{-0.5cm} 10^5}$ 45 \\
1.0 & 71  
    & 10  
    & 195 $\longrightarrow^{\hspace*{-0.5cm} 10^4}$ 65 \\
0.75 & 71  
     & 8  
    & 198 $\longrightarrow^{\hspace*{-0.5cm} 10^4}$ 65 \\
\hline
\hline
\end  {tabular}
\caption{Simulation conditions of the Ising model 
when $\xi_\mathrm{sg}$ is estimated.
Numbers above right arrows denote a time step when the value changes.
}
\label{tab:isg1}
\begin{tabular}{ c c c||c c c }
\hline
\hline
$T$ & $L$ & $N_\mathrm{s}$ &  $T$ & $L$ & $N_\mathrm{s}$ \\
\hline
1.54 & 43 &  2321 & 1.44   & 55 &  1017 \\
1.5  & 43 &  2078 &  1.425   & 59 &  1101 \\
1.475  & 43 &  2208 & 1.42    & 59 &  1288 \\
1.46   & 43 &  1930 & 1.4     & 59 &   809 \\
1.45   & 43 &  1874 & 1.375   & 71 &   412 \\
\hline
\end  {tabular}
\begin{tabular}{ c c c }
\hline
$T$ & $L$ & $N_\mathrm{s}$  \\
\hline
1.35    & 83 &   388 $\longrightarrow^{\hspace*{-0.5cm} 3\cdot 10^5}$ 164\\
1.338   & 83 &   351 $\longrightarrow^{\hspace*{-0.5cm} 3\cdot 10^5}$ 150\\
1.325   & 83 &   306 $\longrightarrow^{\hspace*{-0.5cm} 3\cdot 10^5}$ 214\\
1.3125  & 91 &   458 $\longrightarrow^{\hspace*{-0.5cm} 2\cdot 10^5}$ 409
                     $\longrightarrow^{\hspace*{-0.5cm} 5\cdot 10^5}$ 340
                     $\longrightarrow^{\hspace*{-0.5cm} 9\cdot 10^5}$ 134\\
1.3     & 83 &   130\\
\hline
\hline
\end  {tabular}
\caption{Simulation conditions of the Ising model when 
$\chi_\mathrm{sg}$ is calculated.
Numbers above right arrows denote a time step when the value changes.
}
\label{tab:isg2}
\end  {table}

\begin{table}
\begin{tabular}{ c c c }
\hline
\hline
$T$ & $L$ &  $N_\mathrm{s}$ \\
\hline
0.6 & 39 & 219  $\longrightarrow^{\hspace*{-0.5cm}10^3}$ 624 
                $\longrightarrow^{\hspace*{-0.5cm} 3\cdot 10^4}$ 405\\
0.55& 55 & 1833 $\longrightarrow^{\hspace*{-0.5cm}10^3}$ 174 \\
0.5 & 55 & 247  $\longrightarrow^{\hspace*{-0.5cm} 5\cdot 10^4}$ 99 \\
0.45& 55 & 2056 $\longrightarrow^{\hspace*{-0.5cm} 10^3}$ 675 
                $\longrightarrow^{\hspace*{-0.5cm} 3\cdot 10^4}$ 308\\
0.43& 55 &  167 $\longrightarrow^{\hspace*{-0.5cm} 10^5}$  76 \\
0.41& 55 &  241 \\
0.375& 45  $\longrightarrow^{\hspace*{-0.5cm} 10^5}$ 55 
     & 440 $\longrightarrow^{\hspace*{-0.5cm} 3\cdot 10^4}$ 191
            $\longrightarrow^{\hspace*{-0.5cm} 10^5}$ 26 \\
\hline
\hline
\end  {tabular}
\caption{Simulation conditions of the {\it XY} model when 
$\xi_\mathrm{sg/cg}$
is estimated.
Numbers above right arrows denote a time step when the value changes.
}
\label{tab:xy1}
\end  {table}
\begin{table}
\begin{tabular}{ c c c||c c c }
\hline
\hline
$T$ & $L$ & $N_\mathrm{s}$ &  $T$ & $L$ & $N_\mathrm{s}$ \\
\hline
0.62   & 39 &  128 & 0.50  & 39 &  485 $\longrightarrow^{\hspace*{-0.5cm} 10^5}$ 42 \\
0.6    & 39 &  415 & 0.49  & 55 &  32 \\
0.56   & 39 &  190 & 0.48  & 55 &  143 $\longrightarrow^{\hspace*{-0.5cm} 10^5}$ 30 \\
0.54   & 39 &  426 & 0.47  & 55 &  101 $\longrightarrow^{\hspace*{-0.5cm} 10^5}$ 54 \\
0.53   & 39 &  402 & 0.45  & 55 &  675 $\longrightarrow^{\hspace*{-0.5cm} 3\cdot 10^4}$ 308\\ 
0.52   & 39 &  289 $\longrightarrow^{\hspace*{-0.5cm} 10^5}$ 78 &
0.43   & 55 &  167 $\longrightarrow^{\hspace*{-0.5cm} 10^5}$ 76  \\
0.51   & 39 &  351 &    &&    \\
\hline
\hline
\end  {tabular}
\caption{Simulation conditions of the {\it XY} model when 
$\chi_\mathrm{sg/cg}$ are calculated.
Numbers above right arrows denote a time step when the value changes.
}
\label{tab:xy2}
\end  {table}

\begin{table}
\begin{tabular}{ c c c c c }
\hline
\hline
$T$ & $L$ &$r_\mathrm{min}^\mathrm{sg}$ &$r_\mathrm{min}^\mathrm{cg}$& $N_\mathrm{s}$ \\
\hline
0.29 & 59 & &  & 
 158 $\longrightarrow^{\hspace*{-0.5cm} 10^4}$ 112
     $\longrightarrow^{\hspace*{-0.5cm} 5\cdot 10^4}$ 103
\\
0.28 & 59 & &  & 
174 $\longrightarrow^{\hspace*{-0.5cm} 10^4}$ 81
\\
0.275 & 59 & & & 116
\\
0.27 & 59 &10 & 4& 
200 $\longrightarrow^{\hspace*{-0.5cm} 2\cdot 10^4}$ 98
\\
0.265 & 59 & & & 172
\\
0.26 & 59 & &  & 
249 $\longrightarrow^{\hspace*{-0.5cm} 2\cdot 10^4}$ 177
     $\longrightarrow^{\hspace*{-0.5cm} 10^5}$ 51
\\
0.255 & 59 & & & 103
\\
0.25 & 59 & & 5& 
213 $\longrightarrow^{\hspace*{-0.5cm} 10^4}$ 147
     $\longrightarrow^{\hspace*{-0.5cm} 10^5}$ 56
\\
0.245 & 59 & &  & 135
     $\longrightarrow^{\hspace*{-0.5cm} 10^5}$  58
\\
0.24 & 59 &10 & 5& 121
     $\longrightarrow^{\hspace*{-0.5cm} 10^5}$ 72
\\
0.238 & 59 &10 & 5& 99
\\
0.23 & 59 &10 & 5& 
162 $\longrightarrow^{\hspace*{-0.5cm} 10^4}$ 67
\\
0.225 & 59 &10 & 5& 
185 $\longrightarrow^{\hspace*{-0.5cm} 10^4}$ 61
\\
0.22 & 59 &10 & 5& 110
     $\longrightarrow^{\hspace*{-0.5cm} 2\cdot 10^4}$ 86
\\
0.215 & 59 &10 & 5& 
185 $\longrightarrow^{\hspace*{-0.5cm} 10^4}$ 67
     $\longrightarrow^{\hspace*{-0.5cm} 10^5}$ 18
     $\longrightarrow^{\hspace*{-0.5cm} 3\cdot 10^5}$ 8
\\
0.21 & 75
     $\longrightarrow^{\hspace*{-0.5cm} 2\cdot 10^4}$ 59
          &10 & 5& 86
     $\longrightarrow^{\hspace*{-0.5cm} 2\cdot 10^4}$ 124
     $\longrightarrow^{\hspace*{-0.5cm} 10^5}$ 88
     $\longrightarrow^{\hspace*{-0.5cm} 5\cdot 10^5}$ 26
\\
0.207--0.1925 & 59 &   &  & 
16
\\
0.18 & 59 &10 & 5& 
182 $\longrightarrow^{\hspace*{-0.5cm} 2\cdot 10^4}$ 121
     $\longrightarrow^{\hspace*{-0.5cm} 10^5}$ 17
\\
0.15 & 59 &8 & 5& 
279 $\longrightarrow^{\hspace*{-0.5cm} 5\cdot 10^4}$ 138
     $\longrightarrow^{\hspace*{-0.5cm} 10^5}$ 18
\\
0.1  & 75
     $\longrightarrow^{\hspace*{-0.5cm} 10^5}$ 59
          & 8 & 5& 24
     $\longrightarrow^{\hspace*{-0.5cm} 10^5}$ 10
\\
\hline
\hline
\end  {tabular}
\caption{Simulation conditions of the Heisenberg model.
Numbers above right arrows denote a time step when the value changes.
}
\label{tab:hsg}
\end  {table}

In the {\it XY} model,
a replica number is 64 when the dynamic correlation length is estimated, and
it is 32 when the susceptibility is estimated.
The minimum distance when estimating the correlation length is
$r_\mathrm{min}^\mathrm{sg}=8$ for the spin glass and 
$r_\mathrm{min}^\mathrm{cg}=5$ for the chiral glass.
Simulation conditions when the dynamic correlation length is estimated are
listed in Tab. \ref{tab:xy1}, and
those when the spin-glass susceptibility is calculated are listed in
Tab. \ref{tab:xy2}.

In the Heisenberg model, a replica number is 32.
Simulation conditions are listed in Tab. \ref{tab:hsg}.

\section{Results of the Ising model}
\label{sec:isingresults}

\subsection{Dynamic correlation length}

\begin{figure}
  \begin{center}
  \resizebox{0.45\textwidth}{!}{\includegraphics{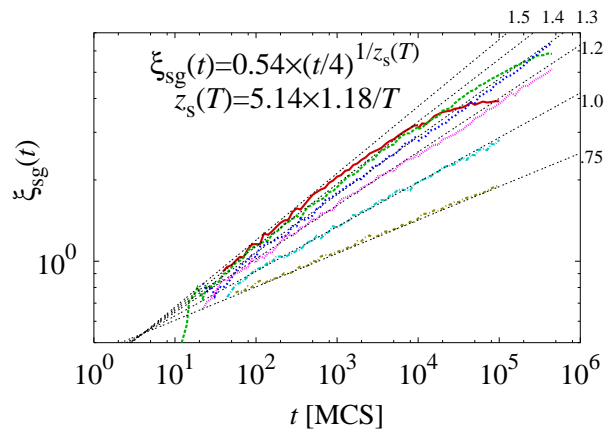}}
  \end{center}
  \caption{
(Color online)
Relaxation functions of the spin-glass correlation length in the Ising model.
The temperatures are denoted aside the fitting lines.
}
\label{fig:xilow1}
\end{figure}

Relaxation functions of the spin-glass correlation length are shown in
Fig. \ref{fig:xilow1}.
Algebraic behavior is observed in the nonequilibrium relaxation process
at all temperatures.
A nonequilibrium dynamic exponent is evaluated from a slope of the relaxation
function.
It is well approximated by
\begin{equation}
\xi_\mathrm{sg}(t)=\xi_0 (t/\tau)^{1/z_\mathrm{s}(T)}
\label{eq:xit}
\end  {equation}
with $\xi_0=0.54$, $\tau=4$, and
\begin{equation}
z_\mathrm{s}(T) = (5.1 \pm 0.1)\times \frac{1.18}{T}.
\end  {equation}
A product, $z_\mathrm{sg}T_\mathrm{sg}$, is a fitting parameter in this 
expression.
It is estimated by the following scaling analysis.

\begin{figure}
  \begin{center}
  \resizebox{0.45\textwidth}{!}{\includegraphics{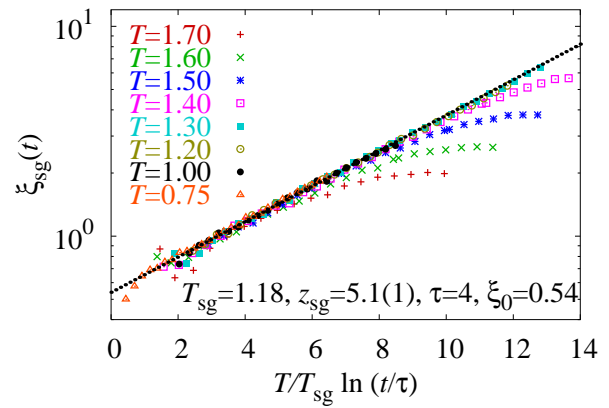}}
  \end{center}
  \caption{
(Color online)
The spin-glass correlation length in the Ising model is plotted against 
$T/T_\mathrm{sg}\ln (t/\tau)$ using $\tau=4$ and $T_\mathrm{sg}=1.18$.
A fitting line with $z_\mathrm{sg}=5.1$ is plotted by
dotted line.
}
\label{fig:xilowsca}
\end{figure}

The relaxation functions of the correlation length are scaled to a single 
line if we plot them against $T/T_\mathrm{sg} \ln(t/\tau)$. 
This plot is proposed 
by Bert {\it et al.}\cite{bert1} for the experimental analysis.
A product $z_\mathrm{sg}T_\mathrm{sg}$ is obtained first from a slope of the
universal scaling line when we plot the data against $T\ln (t/\tau)$.
Here, only a characteristic time $\tau$ is the scaling parameter,
which is very small in the Ising model: $\tau=4$.
Using a value of $z_\mathrm{sg}T_\mathrm{sg}$ we perform the 
finite-time-scaling analysis in the next subsection.
Then a value of $T_\mathrm{sg}=1.18$ is obtained.
Using the estimated critical temperature
we obtain the dynamic critical exponent at the transition temperature,
$z_\mathrm{sg}=(5.1\pm 0.1)$.
The result is shown in Fig. \ref{fig:xilowsca}.
The relaxation functions in the paramagnetic phase show 
a convergence crossover.
It appears later as the temperature approaches $T_\mathrm{sg}$.
We only consider the nonequilibrium process before the crossover time when
evaluating the nonequilibrium dynamic exponent.
A temperature dependence of $z_\mathrm{s}(T)$ is smooth 
above and below the critical temperature. 
This plot suggests that the nonequilibrium process is independent
of the critical temperature.

\subsection{
Finite-time-scaling analysis of $\chi_\mathrm{sg}$
}

\begin{figure}
  \resizebox{0.45\textwidth}{!}{\includegraphics{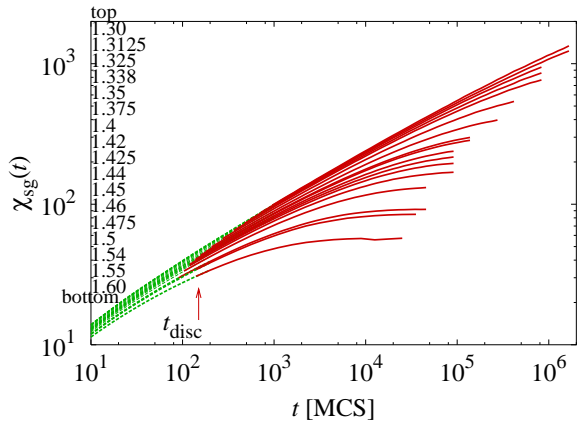}}
  \caption{
(Color online)
Relaxation functions of the spin-glass susceptibility in the Ising model.
Relaxation data after $t_\mathrm{disc}$ (red curves) 
are used in the scaling analysis.
}
\label{fig:Isgchinama}
\end{figure}

The relaxation functions of the spin-glass susceptibility 
$\chi_\mathrm{sg}(t)$ are plotted in Fig.~\ref{fig:Isgchinama}.
We perform the finite-time scaling analysis using these data.
We discard data in the initial relaxation regime before the 
$t_\mathrm{disc}$ step, which are depicted in Fig. \ref{fig:Isgchinama}.
A typical value for $t_\mathrm{disc}$ varies between 100 and 1000.
The initial relaxation is very short compared to the equilibrium 
relaxation.

\begin{figure}
  \begin{center}
  \resizebox{0.45\textwidth}{!}{\includegraphics{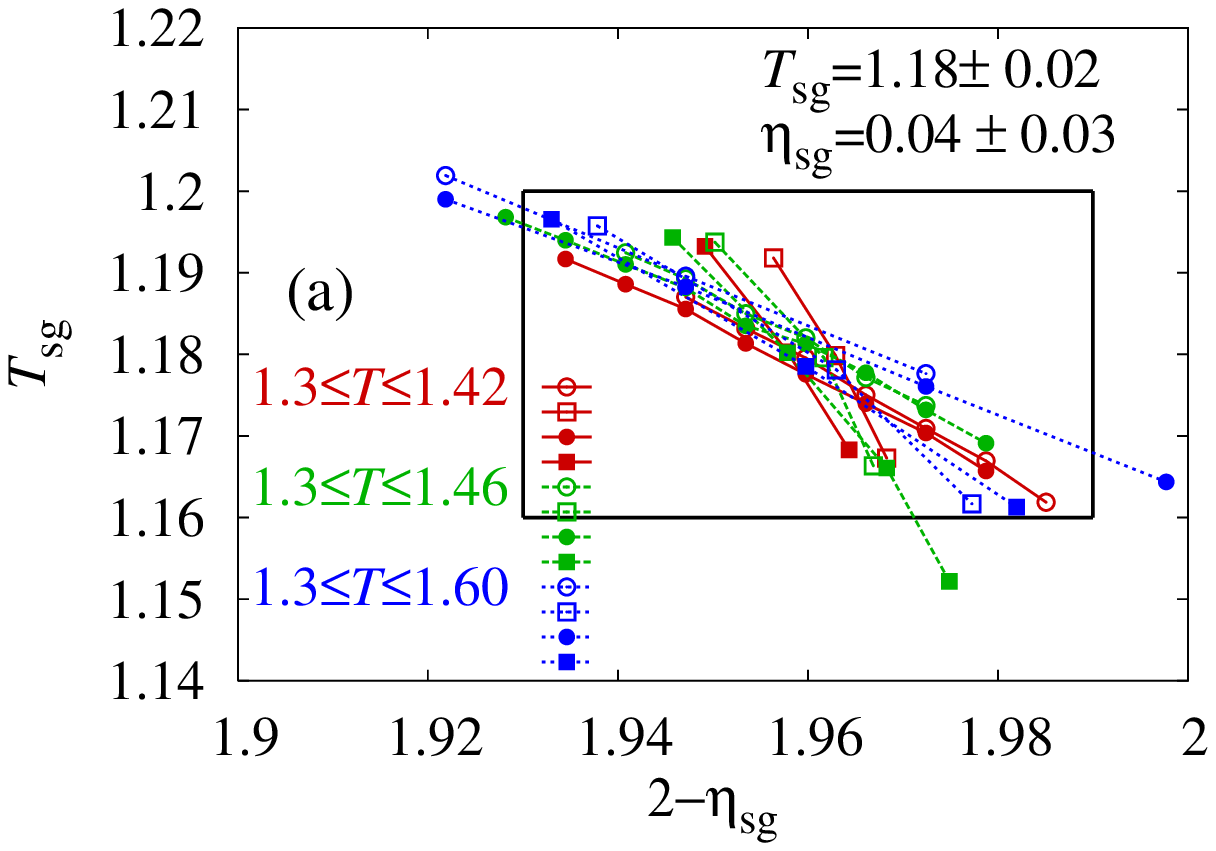}}
  \resizebox{0.45\textwidth}{!}{\includegraphics{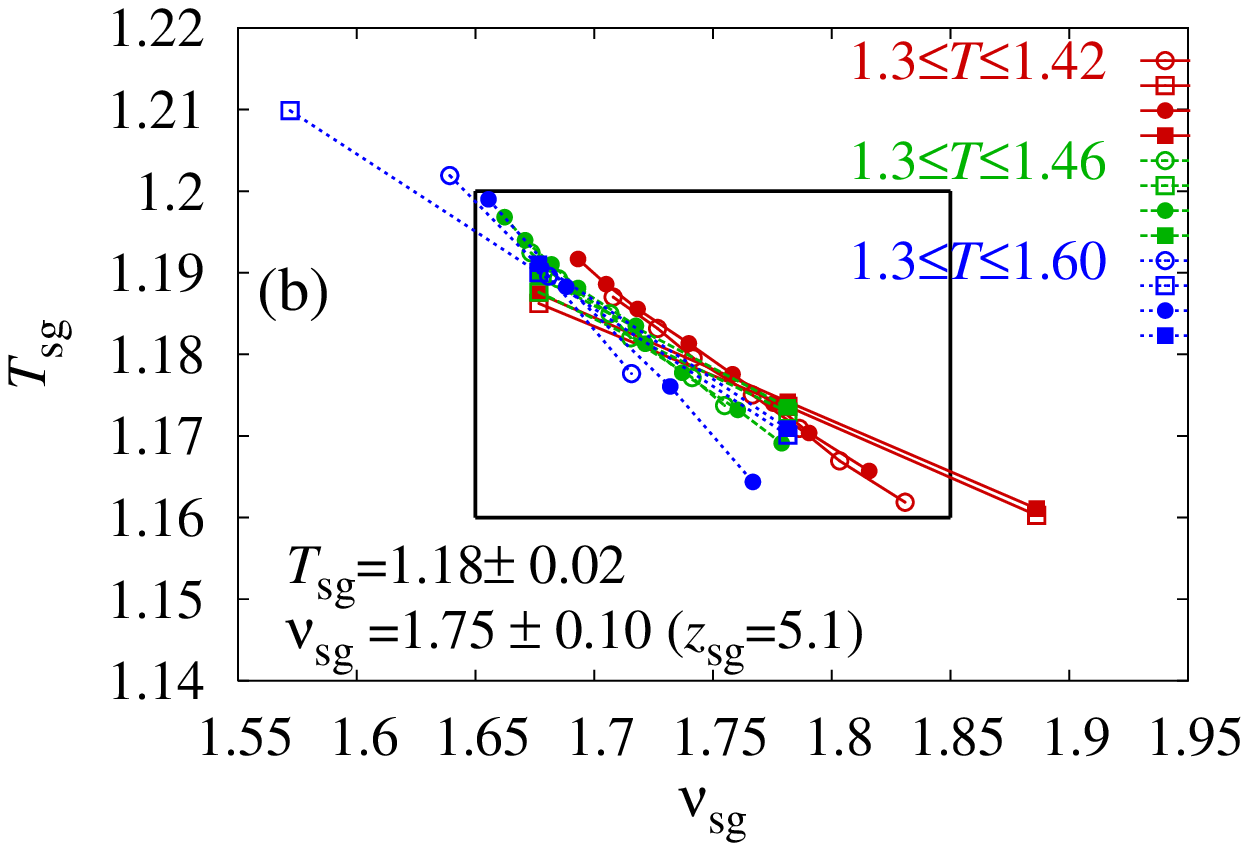}}
  \end{center}
  \caption{
(Color online)
Estimates of the finite-time-scaling parameters in the Ising model.
(a) $T_\mathrm{sg}$ versus $2-\eta_\mathrm{sg}$.
(b) $T_\mathrm{sg}$ versus $\nu_\mathrm{sg}$.
For each temperature range, we plot results of two different procedures 
(circle and square symbols)
and two different discarding steps (filled and unfilled symbols).
Rectangles depict error ranges, which covers most of possible 
scaling parameter sets.
}
\label{fig:isgscale}
\end{figure}
\begin{figure}
  \begin{center}
  \resizebox{0.45\textwidth}{!}{\includegraphics{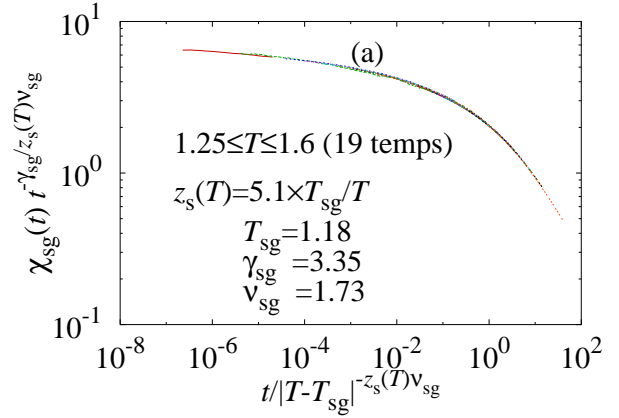}}
  \resizebox{0.45\textwidth}{!}{\includegraphics{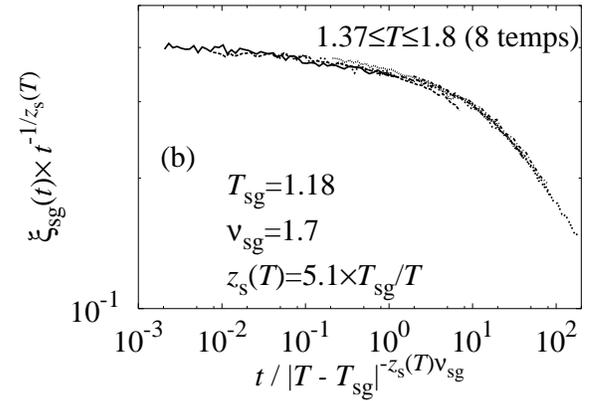}}
  \end{center}
  \caption{
(Color online)
Finite-time-scaling plots in the Ising model.
(a) A scaling plot of $\chi_\mathrm{sg}$ using the obtained results.
(b) A scaling plot of the correlation length in the equilibrium convergence.
}
\label{fig:isgscaleplot}
\end{figure}
We follow the procedures explained in Sec.~\ref{sec:scalingprocedure}.
Three different temperature ranges are considered in this analysis.
One consists of all the 17 temperatures of Fig.~\ref{fig:Isgchinama}
ranging $1.3\le T \le 1.6$.
The second one consists of the lower 12 temperatures 
ranging $1.3\le T \le 1.46$.
The third one consists of the lower 8 temperatures 
ranging $1.3\le T \le 1.42$.
For each temperature range, we adopt three procedures of estimating the 
scaling parameters.
We find that the procedure-2 and the procedure-3 yield consistent results.
Therefore, we compare results of procedure-1 and procedure-2.
We also change the discarding time step, which is 
distinguished in Fig.~\ref{fig:isgscale} with filled and unfilled symbols.
Unfilled symbols depict data when the first $t_\mathrm{disc}$ steps are
discarded, where $t_\mathrm{disc}$ are as depicted in Fig.~\ref{fig:Isgchinama}.
Filled symbols depict data of discarding the first $3t_\mathrm{disc}$ steps.

Figure~\ref{fig:isgscale}(a) shows the obtained critical temperature
plotted versus $2-\eta_\mathrm{sg}$.
Behaviors of the estimated critical temperature depends on 
the procedures, the temperature ranges, and the discarding steps.
These data seem to cross or merge at the most probable parameter point of 
$(2-\eta_\mathrm{sg}, T_\mathrm{sg})\simeq (1.96, 1.18)$.
We put error bars 
so that they cover most of possible scaling parameters.
Figure~\ref{fig:isgscale}(b) shows the critical temperature
plotted versus the critical exponent $\nu_\mathrm{sg}$.
The scaling results cross or merge at 
$(\nu_\mathrm{sg}, T_\mathrm{sg}) \simeq (1.73, 1.18)$.
Two estimates for the transition temperature are consistent.
An exponent $\gamma_\mathrm{sg}$ is estimated from $\eta_\mathrm{sg}$
and $\nu_\mathrm{sg}$.
Using the estimated scaling parameters we plot the scaling functions
in Fig.~\ref{fig:isgscaleplot}(a).
We also plot in Fig.~\ref{fig:isgscaleplot}(b) 
a result of the finite-time scaling analysis of the dynamic correlation length.
It yields the following scaling relation:
\begin{equation}
\xi_\mathrm{sg}(t) t^{-1/z_\mathrm{s}(T)} =  
  \tilde{\xi}_\mathrm{sg}(t/
(T-T_\mathrm{sg})^{-z_\mathrm{s}(T)\nu_\mathrm{sg}})
\label{equ:xisg_scl}
\end  {equation}
The shape of the universal scaling function is robust against changes of
discarding steps and the temperature ranges.

From the present scaling analysis, we conclude the critical temperature and
the critical exponents of the Ising model in three dimensions:
\begin{eqnarray}
T_\mathrm{sg} &=& 1.18 \pm 0.02\\
\gamma_\mathrm{sg}&=& 3.4 \pm 0.2\\
\nu_\mathrm{sg} &=& 1.75 \pm 0.10\\
\eta_\mathrm{sg}&=&0.04\pm 0.03\\
z_\mathrm{sg}&=&5.1\pm 0.1.
\end  {eqnarray}

Previous estimates
\cite{Bhatt,Ogielski,KawashimaY,palassini,ballesteros,maricampbell}
for $T_\mathrm{sg}$ lie between 1.1 and 1.2.
They are roughly categorized into two groups.
Group I: $T_\mathrm{sg}$ close to 1.1 
and $\nu_\mathrm{sg}$ close to 2.\cite{KawashimaY,ballesteros}
Group II: $T_\mathrm{sg}$ close to 1.2 and $\nu_\mathrm{sg}$ 
close to 1.3.\cite{Bhatt,Ogielski,maricampbell}
Our estimate for $T_\mathrm{sg}$ belongs to the latter group.
On the other hand,
an estimate for $\nu_\mathrm{sg}$ is rather close to the former group and
agrees with the experimental result,\cite{Aruga}
which is $\nu_\mathrm{sg}\sim 1.7$.
It is very hard to determine a value of $\nu_\mathrm{sg}$.

Recently, Campbell {\it et. al.} \cite{campbell-huku-taka}
claimed that a choice of a scaling variable may affect the scaling results.
Particularly, a value of $\nu$ changes.
We applied their method and checked the claim.
A scaling plot is possible with $T_\mathrm{sg}=1.18$, $\nu_\mathrm{sg}=2.18$,
and $\gamma_\mathrm{sg}=4.15$,
when we use a scaling vaiable $(\beta_\mathrm{sg}-\beta)$ instead
of $(T-T_\mathrm{sg})$.
The critical temperature and the anomalous exponenet $\eta_\mathrm{sg}$ are
not affected by this change.
We consider that these two estimates are reliable.
To the contrary,
two exponents, $\nu_\mathrm{sg}$ and $\gamma_\mathrm{sg}$, become larger than
our original estimates and are consistent with the values of group I.
We cannot exclude out either estimate within our amount of simulations.

\section{Results of the {\it XY} model}
\label{sec:xyresults}

\subsection{Brief introduction to the vector models}

It is widely accepted that the spin-glass transition occurs in the Ising
model.\cite{Bhatt,Ogielski,KawashimaY,palassini,ballesteros,maricampbell}
However, 
spins of many spin-glass materials are of the Heisenberg type.
Early numerical investigations \cite{McMillan,Morris,Olive}
concluded that there is no spin-glass transition in this model.
Kawamura \cite{KawamuraXY1,KawamuraXY2,KawamuraXY3,KawamuraH1}
proposed the chirality-mechanism 
in order to explain the real spin-glass transition.
The spin-glass transitions in real materials occur
driven by the chiral-glass transition.
This scenario assumes that there is no spin-glass transition in the
isotropic model.\cite{HukushimaH,HukushimaH2}
Recently, there are several investigations which suggest the simultaneous
transition.\cite{Matsubara3,Matsubara4,Endoh2,nakamura,Lee}

The situation is same in the {\it XY} spin-glass model in three dimensions.
A domain-wall energy analysis \cite{Morris} and a Monte Carlo simulation
analysis \cite{Jain} found no spin-glass transition.
However, a possibility of a finite spin-glass transition has been suggested
by recent investigations \cite{Lee,Maucourt,Kosterlitz,Granato,yamamoto1}.
Since the chirality is defined by spin variables, the chiral-glass
transition occurs trivially if the spin-glass transition occurs.
Therefore, it is crucial to check whether or not both transitions occur at the
same temperature.
This is a very naive problem.
For example, the spin-chirality separation is observed in 
the fully frustrated nonrandom {\it XY} model in two dimensions.\cite{OzIto2}
However, the difference between the critical temperatures is only 1\%.

\subsection{Dynamic correlation length}

\begin{figure}
  \begin{center}
  \resizebox{0.45\textwidth}{!}{\includegraphics{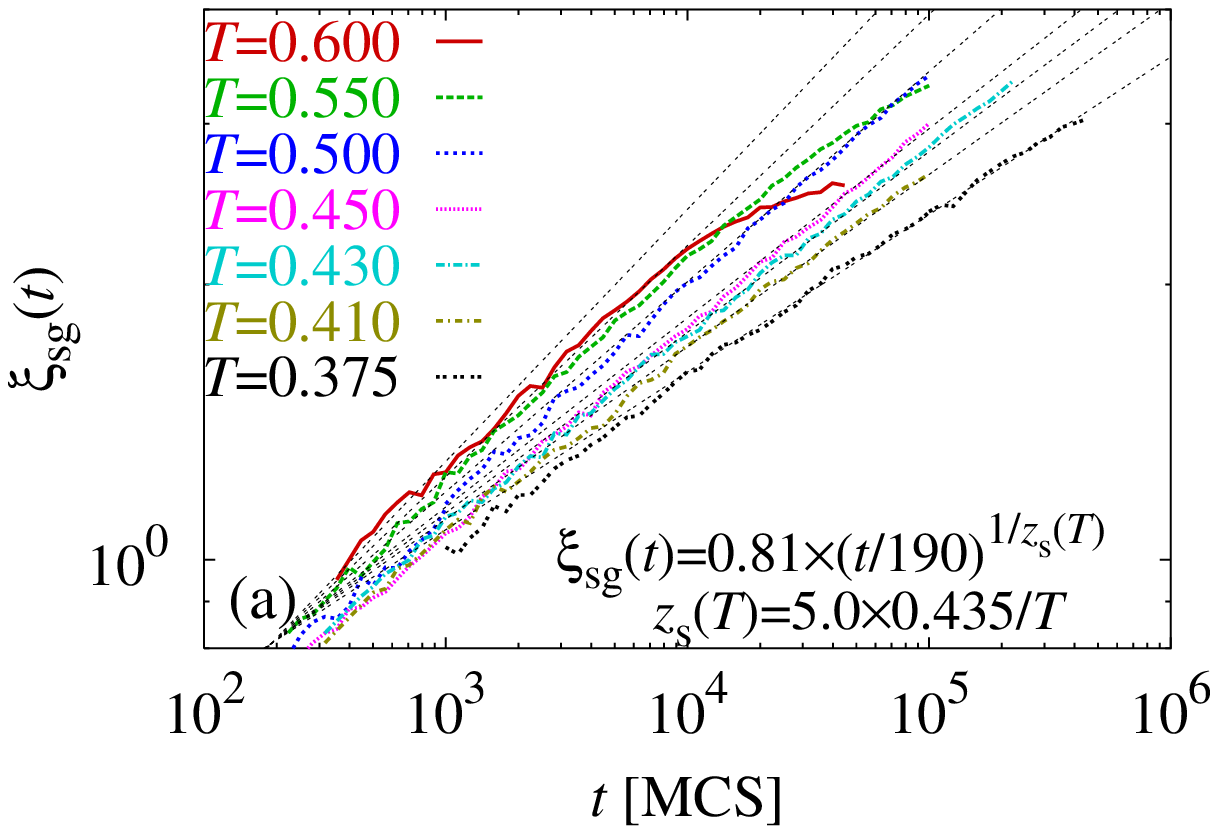}}
  \resizebox{0.45\textwidth}{!}{\includegraphics{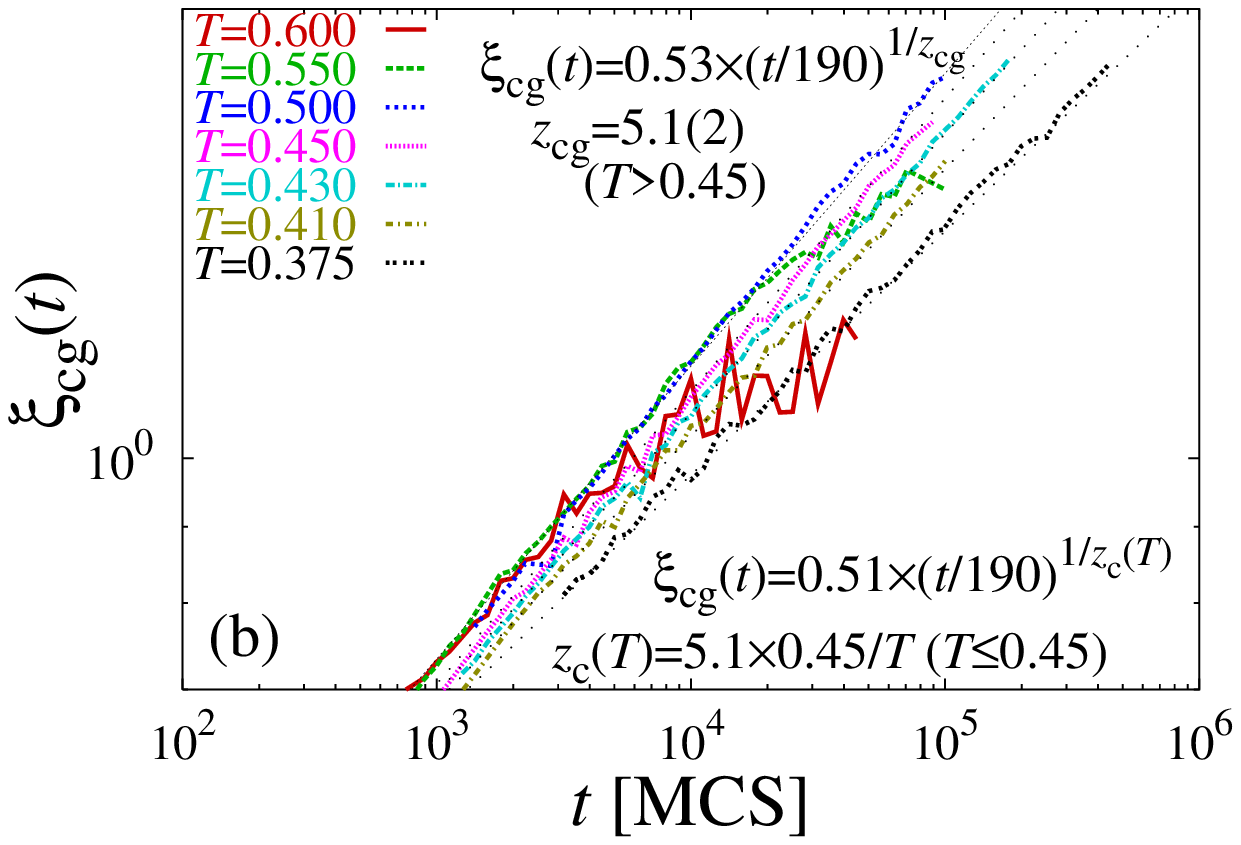}}
  \resizebox{0.45\textwidth}{!}{\includegraphics{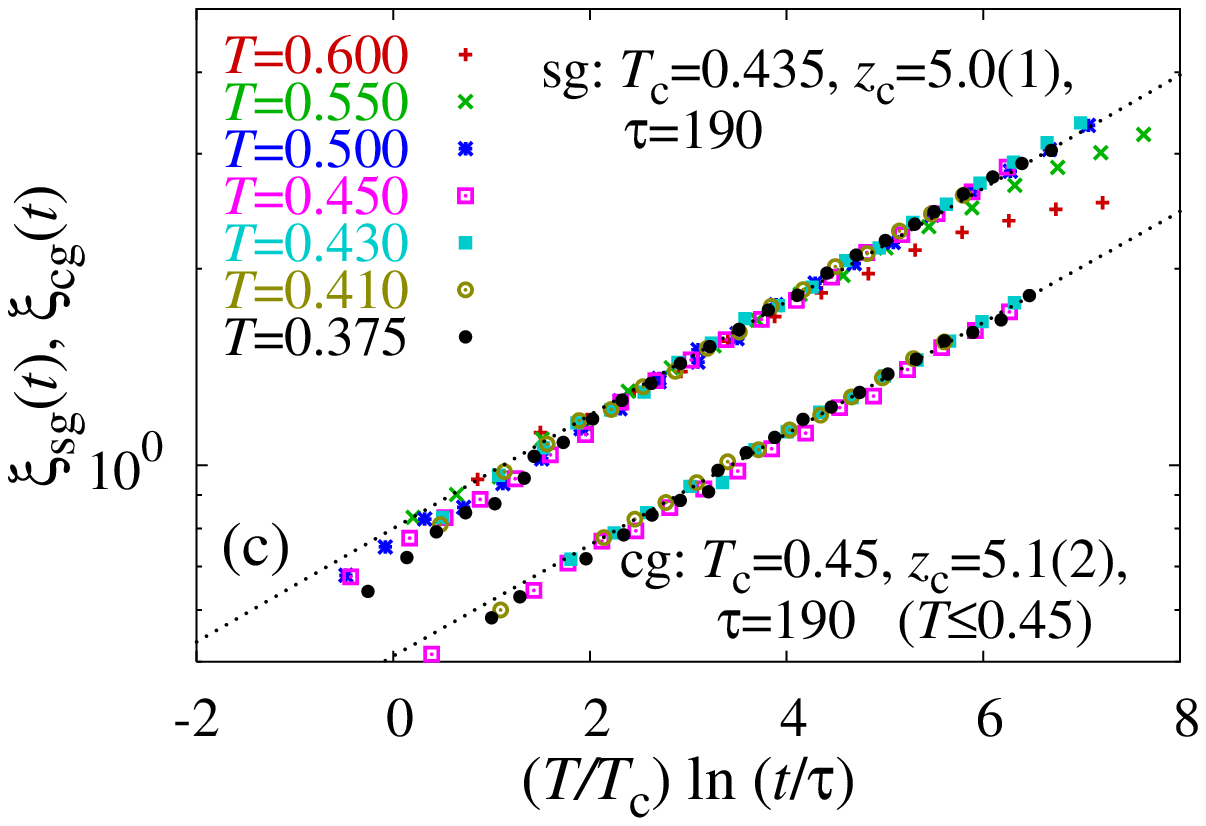}}
  \end{center}
  \caption{
(Color online)
Relaxation functions of the correlation lengths in the {\it XY} model.
(a) The spin-glass correlation length.
(b) The chiral-glass correlation length.
(c) Scaling plots of $\xi_\mathrm{sg}$ and $\xi_\mathrm{cg}$ 
versus $T/T_\mathrm{c}\ln (t/\tau)$. 
Scaling parameters are as denoted in the figure.
The chiral-glass scaling is only possible at $T\le 0.45$.
}
\label{fig:xi-xy}
\end{figure}

Relaxation functions of the spin- and the chiral-glass correlation length 
are shown in Figs.~\ref{fig:xi-xy}(a) and (b).
Algebraic behaviors are observed in the nonequilibrium relaxation process 
both above and below the critical temperature.
Slope of relaxation functions of the spin-glass correlation length 
continuously depends on the temperature.
There is no anomaly at the transition temperature, $T_\mathrm{sg}=0.435$,
which will be estimated in the next subsection.
This is the same behavior with the Ising model.
On the other hand, 
relaxation functions of the chiral-glass correlation length do not depend on
the temperature at higher temperatures ($T\ge 0.45$).
It changes the behavior at $T\simeq 0.45$ becoming temperature-dependent at 
lower temperatures.
This is a distinct difference between the spin- and the chiral-glass 
nonequilibrium dynamics.
From the scaling analysis of the correlation lengths,
we obtain the temperature dependences of the nonequilibrium dynamic exponents.
Figure \ref{fig:xi-xy}(c) shows results of the scaling.
The scaling is possible both below and above $T_\mathrm{sg}$ for the
spin-glass correlation length, but it is only possible below 
$T_\mathrm{sg}\simeq T_\mathrm{cg}=0.45$ for the chiral-glass one.
It is also noted that a characteristic time is common between the
spin and the chirality, and it is larger than the Ising one.

We obtain temperature dependences of the dynamic exponent
for the spin glass ($z_\mathrm{s}(T)$) and 
the chiral glass ($z_\mathrm{c}(T)$) as follow.
\begin{eqnarray}
z_\mathrm{s}(T)&=& (5.0\pm 0.1) \times \frac{0.435}{T}
\label{eq:xyzTsg}
\\
z_\mathrm{c}(T)&=& (5.1\pm 0.2)  ~~~~~~~~~~~~~(T> 0.45)
\label{eq:xyzTcg}
\\
z_\mathrm{c}(T)&=& (5.1\pm 0.2) \times \frac{0.45}{T} ~~~(T\le 0.45)
\end  {eqnarray}
Here, we use $T_\mathrm{sg}=0.435$ and $T_\mathrm{cg}=0.45$, which will be
estimated in the next subsection.
A coefficient to the temperature-ratio term 
is the dynamic critical exponent at the transition temperature.
It agrees with each other within error bars.
Therefore, the dynamic critical exponents of the spin-glass and the 
chiral-glass transitions are considered to take the same value.
The nonequilibrium dynamic exponent of the chiral glass changes behavior 
when the spin-glass exponent equals to the the chiral-glass one.
The spin-glass transition occurs at this temperature.
Both exponents behave in the same manner below the temperature.

\subsection{
Finite-time-scaling analysis of $\chi_\mathrm{sg}$ and $\chi_\mathrm{cg}$
}

\begin{figure}
  \begin{center}
  \resizebox{0.45\textwidth}{!}{\includegraphics{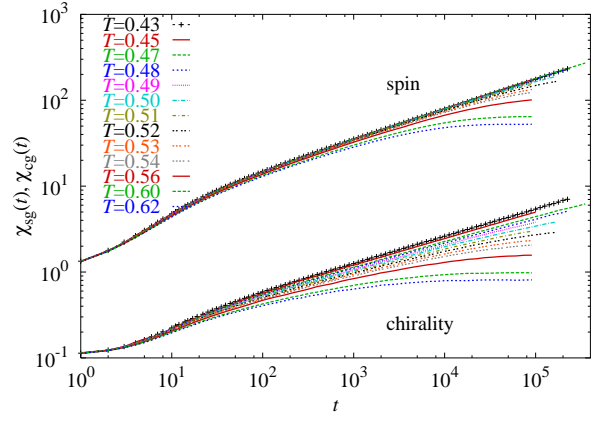}}
  \end{center}
  \caption{
(Color online)
Relaxation functions of the spin- and the chiral-glass susceptibility in
the {\it XY} model.
}
\label{fig:xychinama}
\end{figure}

Figure \ref{fig:xychinama} shows the relaxation functions of the 
spin- and the chiral-glass susceptibility.
Some of the data are taken from our previous work.\cite{yamamoto1}
We added several temperature points and 
increased a total Monte Carlo step
particularly near the transition temperature.
In the previous work, we supposed that the dynamic exponent is
independent from the temperature.
Crossing behaviors of the scaling results were not good.
Consequently, the error bars of obtained results became large.
These failures are possibly due to the temperature dependence of 
the nonequilibrium dynamic exponent.
In the present analysis we use Eqs.~(\ref{eq:xyzTsg}) and (\ref{eq:xyzTcg}).
A scaling procedure is same as the Ising case.

\begin{figure}
  \begin{center}
  \resizebox{0.45\textwidth}{!}{\includegraphics{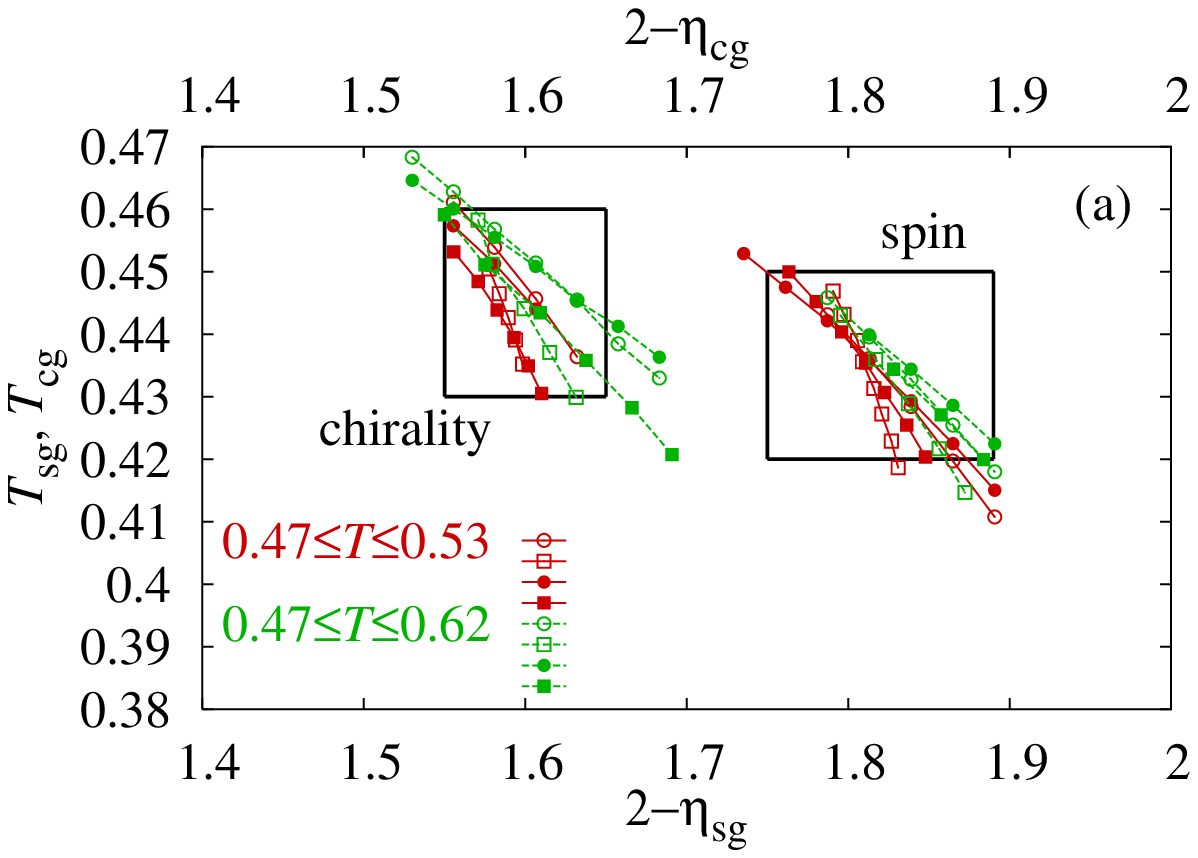}}
  \resizebox{0.45\textwidth}{!}{\includegraphics{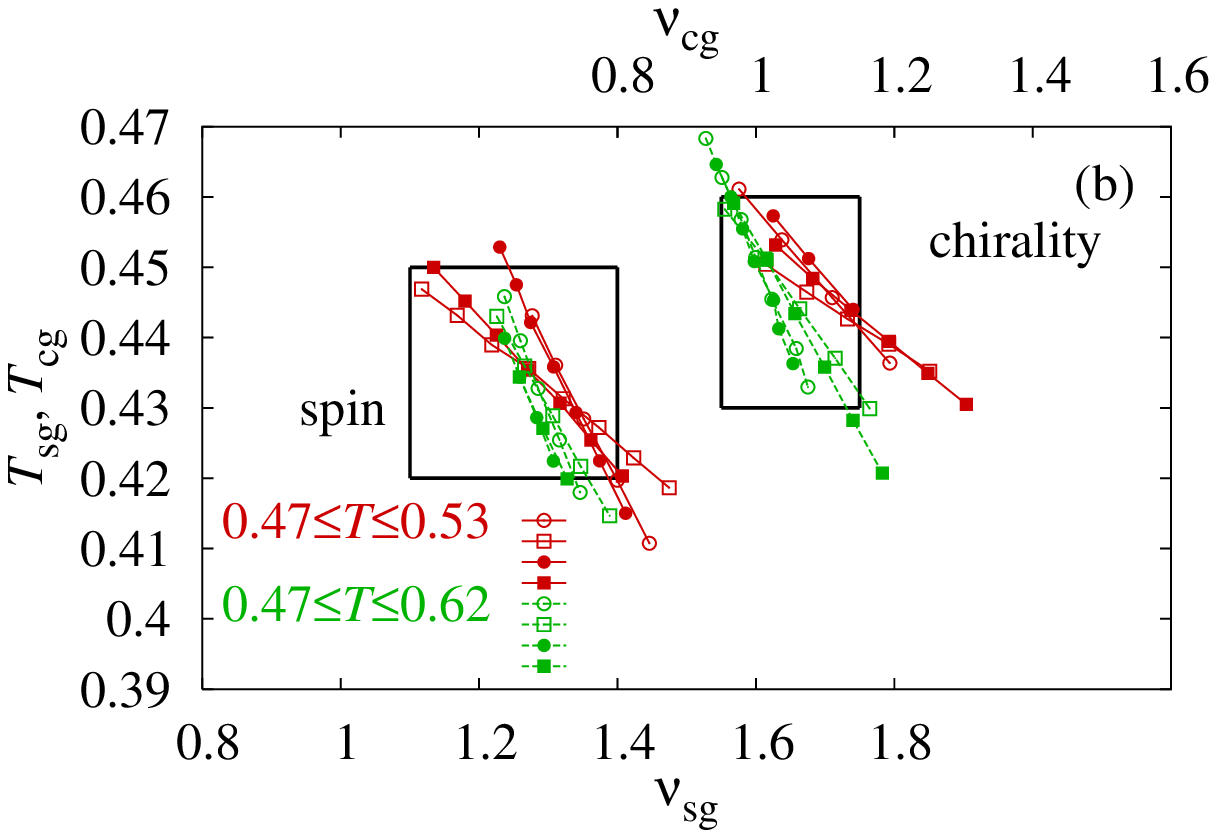}}
  \end{center}
  \caption{
(Color online)
Estimations of the finite-time-scaling parameters in the {\it XY} model.
(a) $T_\mathrm{c}$ versus $2-\eta$.
(b) $T_\mathrm{c}$ versus $\nu$.
Spin-glass(chiral-glass) results are plotted against the lower(upper) x-axis.
The circle symbols are results of the scaling procedure-1, and
the square symbols are those of the procedure-2.
Unfilled symbols depict data when the first 400 steps are discarded.
Filled symbols depict data when the first 1000 steps are discarded.
}
\label{fig:scale-xy}
\end{figure}
\begin{figure}
  \begin{center}
  \resizebox{0.45\textwidth}{!}{\includegraphics{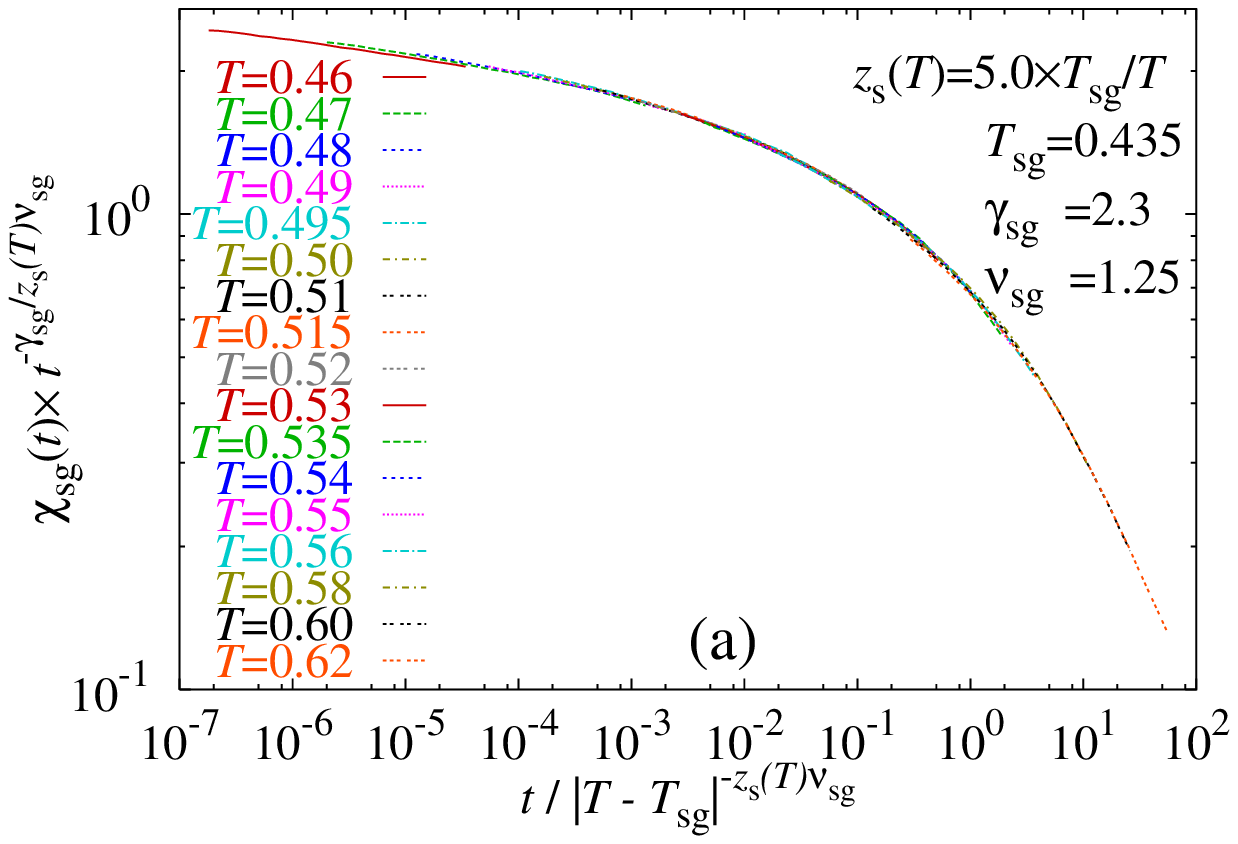}}
  \resizebox{0.45\textwidth}{!}{\includegraphics{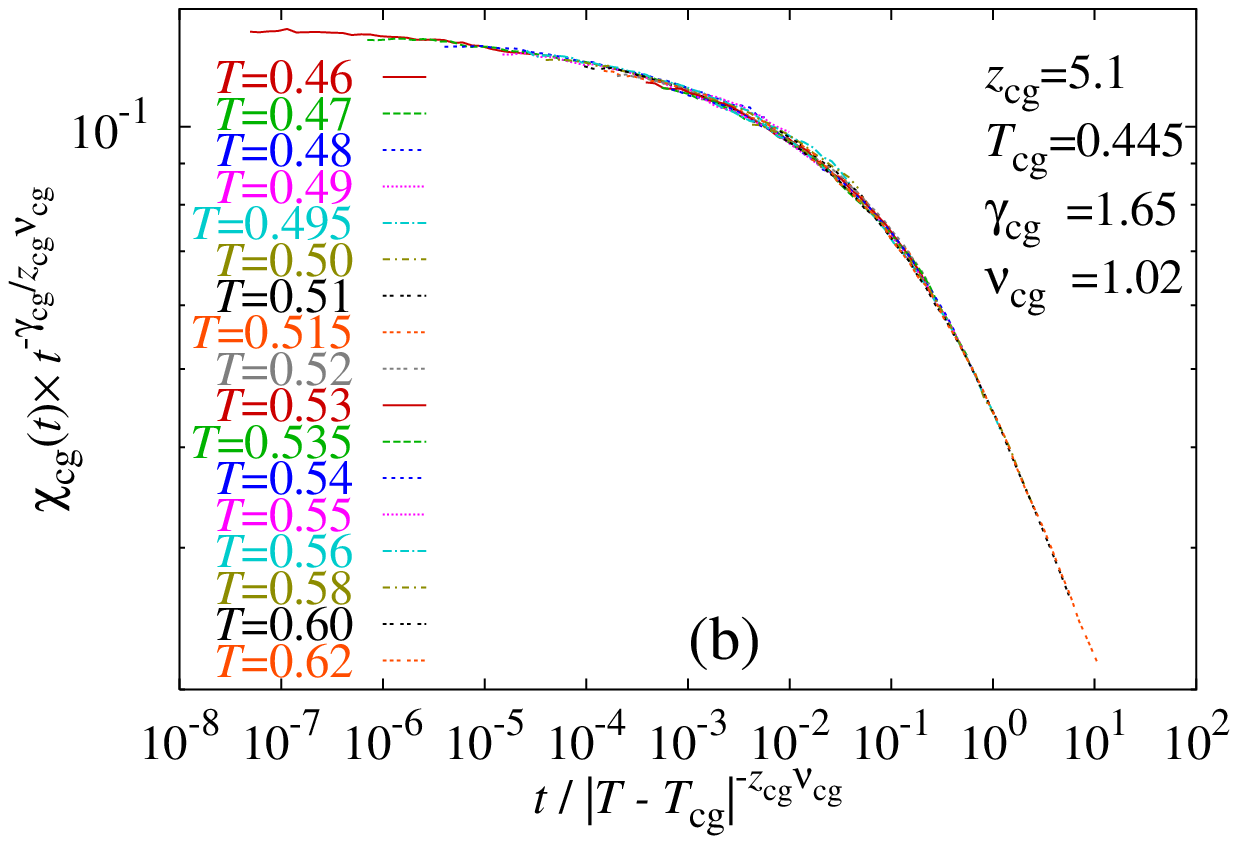}}
  \end{center}
  \caption{
(Color online)
Finite-time-scaling plots of
(a) $\chi_\mathrm{sg}$ and
(b) $\chi_\mathrm{cg}$
using the estimated scaling 
parameters in the {\it XY} model.
}
\label{fig:scale-xyplot}
\end{figure}

Figures \ref{fig:scale-xy} (a) and (b) show estimations of the 
scaling parameters.
We determine the best estimates from the figure so that the results are robust
against the difference of procedures.
For the spin-glass transition:
\begin{eqnarray}
T_\mathrm{sg} &=& 0.435 \pm 0.015\\
\gamma_\mathrm{sg}&=& 2.30 \pm 0.25\\
\nu_\mathrm{sg} &=& 1.25 \pm 0.15\\
\eta_\mathrm{sg} &=& 0.18 \pm 0.07\\
z_\mathrm{sg} &=& 5.0 \pm 0.1.
\end{eqnarray}
For the chiral-glass transition:
\begin{eqnarray}
T_\mathrm{cg} &=& 0.445 \pm 0.015\\
\gamma_\mathrm{cg}&=& 1.65 \pm 0.20\\
\nu_\mathrm{cg} &=& 1.05 \pm 0.10\\
\eta_\mathrm{cg} &=& 0.40 \pm 0.05\\
z_\mathrm{cg} &=& 5.1 \pm 0.2.
\end{eqnarray}
The spin- and the chiral-glass transition temperatures cannot be
distinguished to be different.
The chiral-glass transition occurs at a little higher temperature but
the difference is within the error bars.
A value of the dynamic critical exponent
coincides with each other within the error bar.
Both transitions are considered to be dynamically universal.
On the other hand, the static exponents take different values.

The scaling plot using the obtained results is shown in 
Figs. \ref{fig:scale-xyplot}(a) and (b).
All data suitably fall onto a single scaling function.
The obtained transition temperature is between
our previous estimate \cite{yamamoto1} supposing a constant dynamic
exponent and other estimates.\cite{KawamuraXY3,Granato}
Our previous estimates are
$T_\mathrm{sg}=0.455(15)$, $\gamma_\mathrm{sg}=1.7(2)$, 
$z_\mathrm{sg}\nu_\mathrm{sg}=4.8(4)$ for the spin-glass transition and 
$T_\mathrm{cg}=0.467(10)$, $\gamma_\mathrm{cg}=1.4(1)$, 
and $z_\mathrm{cg}\nu_\mathrm{cg}=4.7(2)$ for the
chiral-glass transition.
The results of the spin-glass transition are strongly influenced by the 
temperature dependence of $z_\mathrm{s}(T)$.
Granato \cite{Granato} provided the following for the spin-glass transition:  
$T_\mathrm{sg}=0.39(2)$, $\nu=1.2(2)$, and $z=4.4(3)$.
Kawamura and Li \cite{KawamuraXY3} provided the following
for the chiral-glass transition:
$T_\mathrm{cg}=0.39(3)$, $\nu=1.2(2)$, 
$\gamma/\nu=1.85(20)$, and $z=7.4(10)$.
These transition temperatures are lower then our present estimates.

\section{Results of the Heisenberg model}
\label{sec:Hsgresults}

\subsection{Dynamic correlation length}

\begin{figure}
  \begin{center}
  \resizebox{0.45\textwidth}{!}{\includegraphics{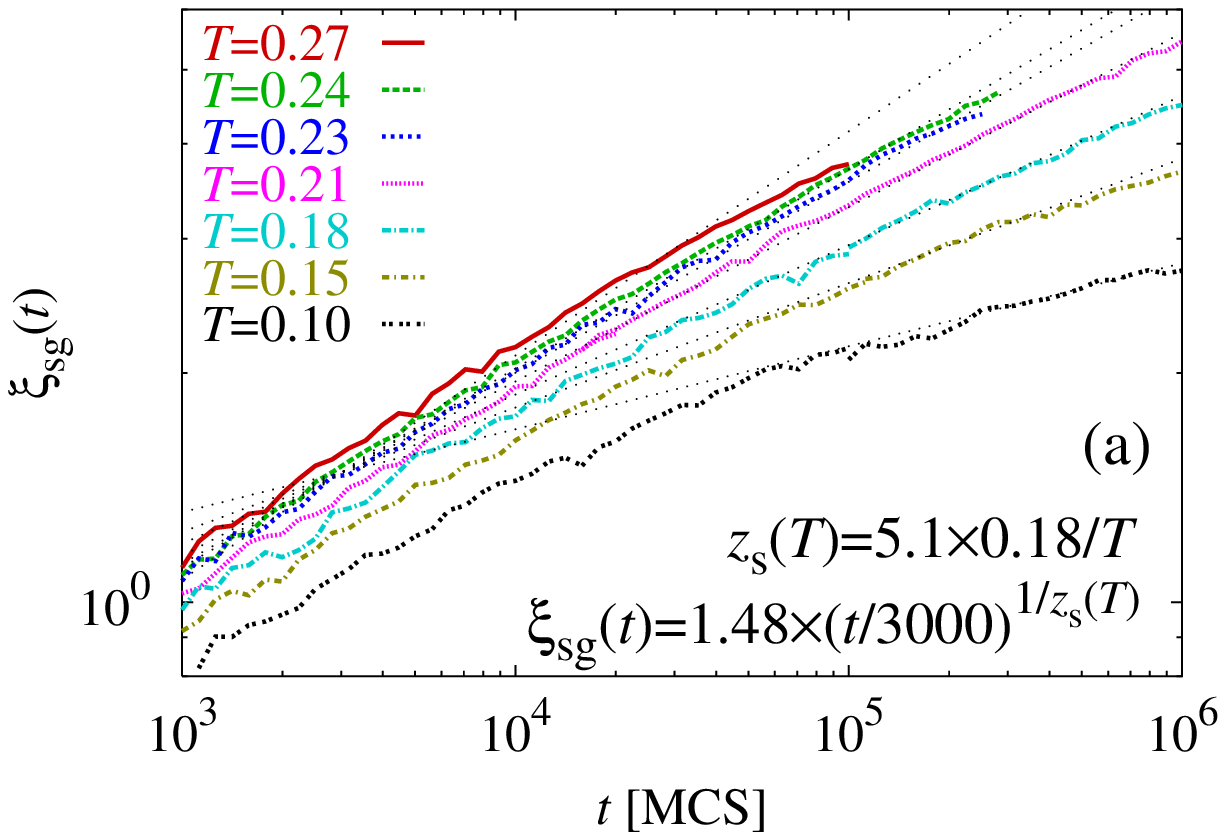}}
  \resizebox{0.45\textwidth}{!}{\includegraphics{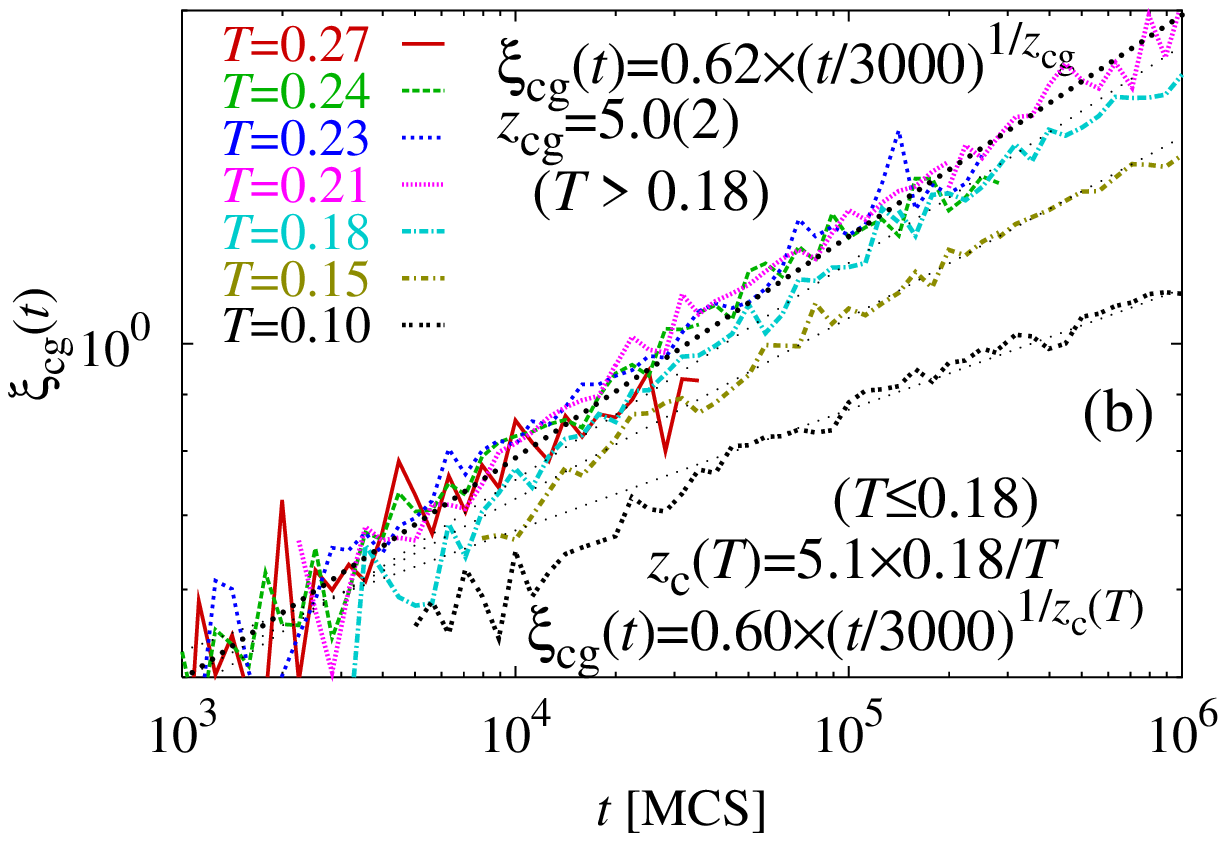}}
  \resizebox{0.45\textwidth}{!}{\includegraphics{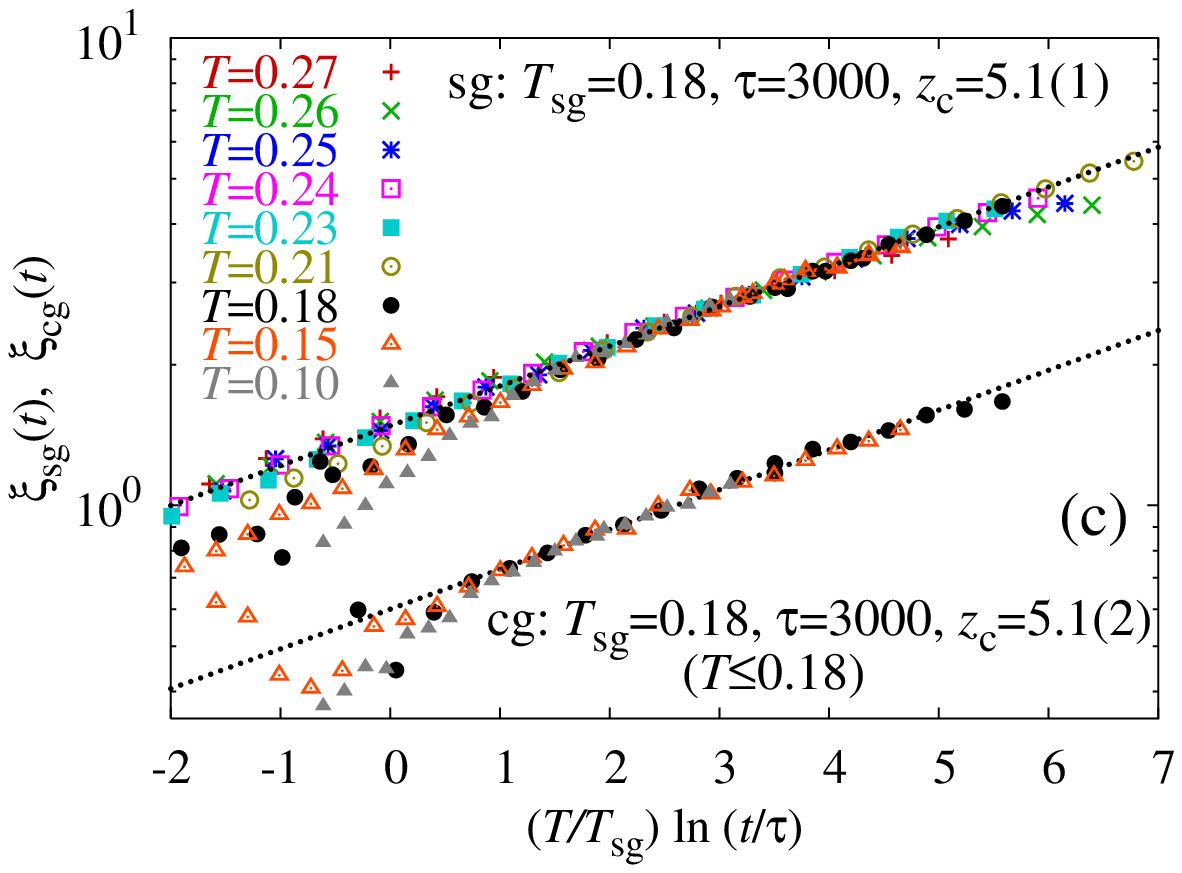}}
  \end{center}
  \caption{
(Color online)
Relaxation functions of the correlation lengths in the Heisenberg model.
(a) The spin-glass correlation length.
(b) The chiral-glass correlation length.
Fitting lines are plotted with broken lines.
(c) The scaling plot of $\xi_\mathrm{sg}$ plotted versus 
$T/T_\mathrm{sg}\ln (t/\tau)$. 
}
\label{fig:xi-Hsg}
\end{figure}

Relaxation functions of the spin- and the chiral-glass correlation lengths
are shown in Figs.~\ref{fig:xi-Hsg}(a) and (b).
We observe algebraic increases of both correlation lengths.
We estimate the dynamic exponent using these data.

The chiral-glass exponent may have a rather large error bar because
numerical fluctuations of the chiral-glass data are large.
Evaluatios of the chiral-glass quantities are generally difficult because 
the chiral-glass correlations are much smaller than the spin-glass ones
as shown in Fig.~\ref{fig:fsgising}(b).
It is necessary to increase a sample number 
in order to suppress the fluctuations.
Within the present computational facilities, it is not realistic to increase
a Monte Carlo step number until the chiral-glass correlation length grows 
comparable to the present spin-glass data.
For example, we need a hundred times longer steps 
to wait until $\xi_\mathrm{cg}$ reaches five lattice spacings.

It is also noted that the algebraic increase in the low-temperature phase
appears later as the temperature decrease.
It is considered to be an outcome of the slow dynamics of the spin-glass phase.
A time range to estimate the dynamic exponent is restricted to be short.
Reliability of the estimated value may become poor.

Even though the numerical fluctuations are large,
we observe a temperature-dependence of the dynamic exponent
as in the Ising model and the {\it XY} model.
Logarithmic slope of the spin-glass correlation length depends on the
temperature irrespective of the transition temperature.
That of the chiral-glass correlation length depends on the temperature
only when $T<0.18$.
These evidences are checked by a scaling plot of Fig.~\ref{fig:xi-Hsg}(c).
The spin-glass data can be scaled to a single line both above and below the
critical temperature, $T_\mathrm{sg}=0.18$, which will be estimated in the
next subsection.
On the other hand, the chiral-glass data are scaled only below $T_\mathrm{sg}$.
The high temperature data systematically deviate from the scaling line.
They are better approximated by a temperature-independent algebraic
function: $\xi_\mathrm{cg}=0.62\times (t/3000)^{1/5.0}$.
Therefore, the chiral-glass dynamic exponent is constant above $T_\mathrm{sg}$
and becomes temperature dependent below $T_\mathrm{sg}$.
The temperature dependence of the chiral-glass dynamic exponent 
below $T_\mathrm{sg}$ is same as the spin-glass one.
These behaviors are same as the {\it XY} model.
A difference is a characteristic time $\tau$, 
which is much larger than the Ising and the {\it XY} models.

The temperature dependences of the nonequilibrium dynamic exponent
in the Heisenberg model,
$z_\mathrm{s}(T)$ for the spin glass $z_\mathrm{c}(T)$ for the
chiral glass, are summarized as follow.
\begin{eqnarray}
z_\mathrm{s}(T)&=& (5.1\pm 0.1) \times \frac{0.18}{T} 
\label{eq:HsgzTsg}
\\
z_\mathrm{c}(T)&=& (5.0\pm 0.2)  ~~~~~~~~~~~~~(T>0.18)
\\
z_\mathrm{c}(T)&=& (5.1\pm 0.2) \times \frac{0.18}{T} ~~~(T\le 0.18)
\label{eq:HsgzTcg}
\end  {eqnarray}
Values of the nonequilibrium dynamic exponent 
at $T=T_\mathrm{sg}=0.18$ agree with each other.
This is the temperature where the chiral-glass
dynamic exponent changes the behavior.
It can be considered that dominant dynamics changes when
the $T$-dependent $z_\mathrm{s}(T)$ catches up with 
the $T$-independent $z_\mathrm{c}(T)$ at $T=0.18$.
The spin-glass {\it nonequilibrium} dynamic exponent, $z_\mathrm{s}(T)$, is 
larger than the chiral-glass dynamic {\it critical} exponent, $z_\mathrm{cg}$,
below the temperature.
Then, the dynamics of the chirality obeys that of the spin because
the chirality is defined by the spin.
As the result, $z_\mathrm{c}(T)$ behaves in a same manner with
$z_\mathrm{s}(T)$.

\subsection{Finite-time-scaling analysis of $\chi_\mathrm{sg}$ 
and $\chi_\mathrm{cg}$}

\begin{figure}
  \begin{center}
  \resizebox{0.45\textwidth}{!}{\includegraphics{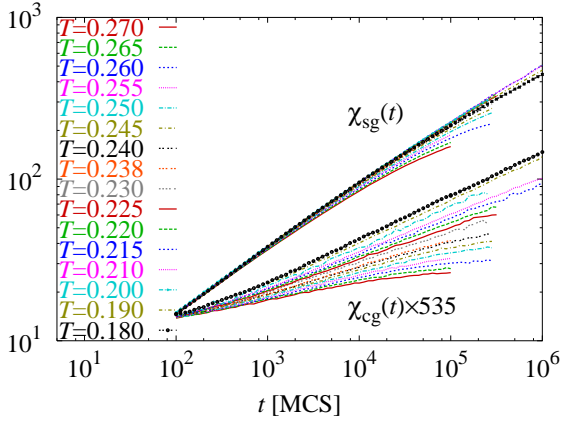}}
  \end{center}
  \caption{
(Color online)
Relaxation functions of the spin-glass and the chiral-glass susceptibility
in the Heisenberg model.
Data of the chiral-glass susceptibility are multiplied by 535 in order to
show in the same scale with the spin-glass susceptibility.
}
\label{fig:chisg-Hsg}
\end{figure}

Figure \ref{fig:chisg-Hsg} shows relaxation functions of the 
spin- and the chiral-glass susceptibility.
Data of the chiral-glass susceptibility are multiplied by 535 in order to
show them in a same scale with the spin-glass susceptibility.
Amplitude of the chiral-glass susceptibility is roughly 1000 times smaller
than the spin-glass one.
The chiral-glass order is very small.

\begin{figure}
  \begin{center}
  \resizebox{0.45\textwidth}{!}{\includegraphics{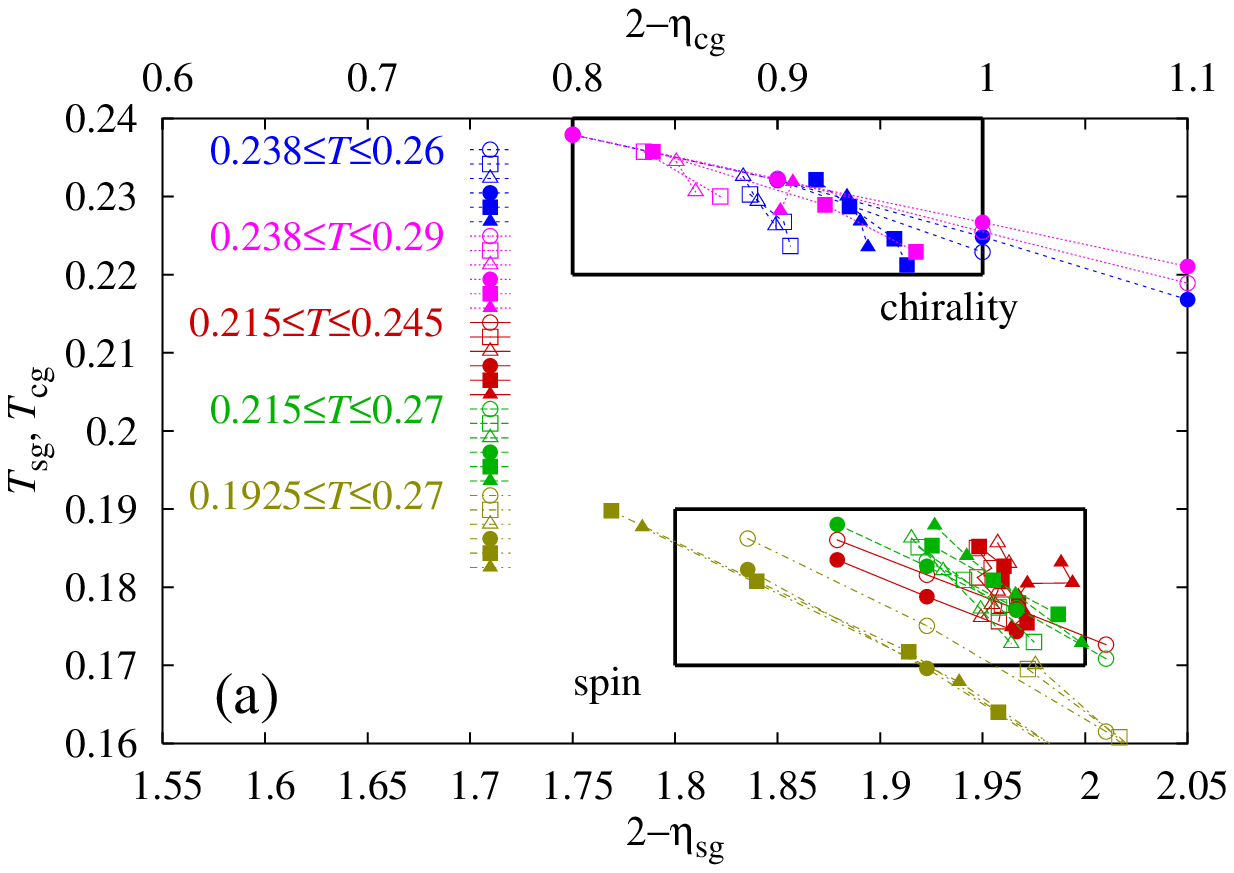}}
  \resizebox{0.45\textwidth}{!}{\includegraphics{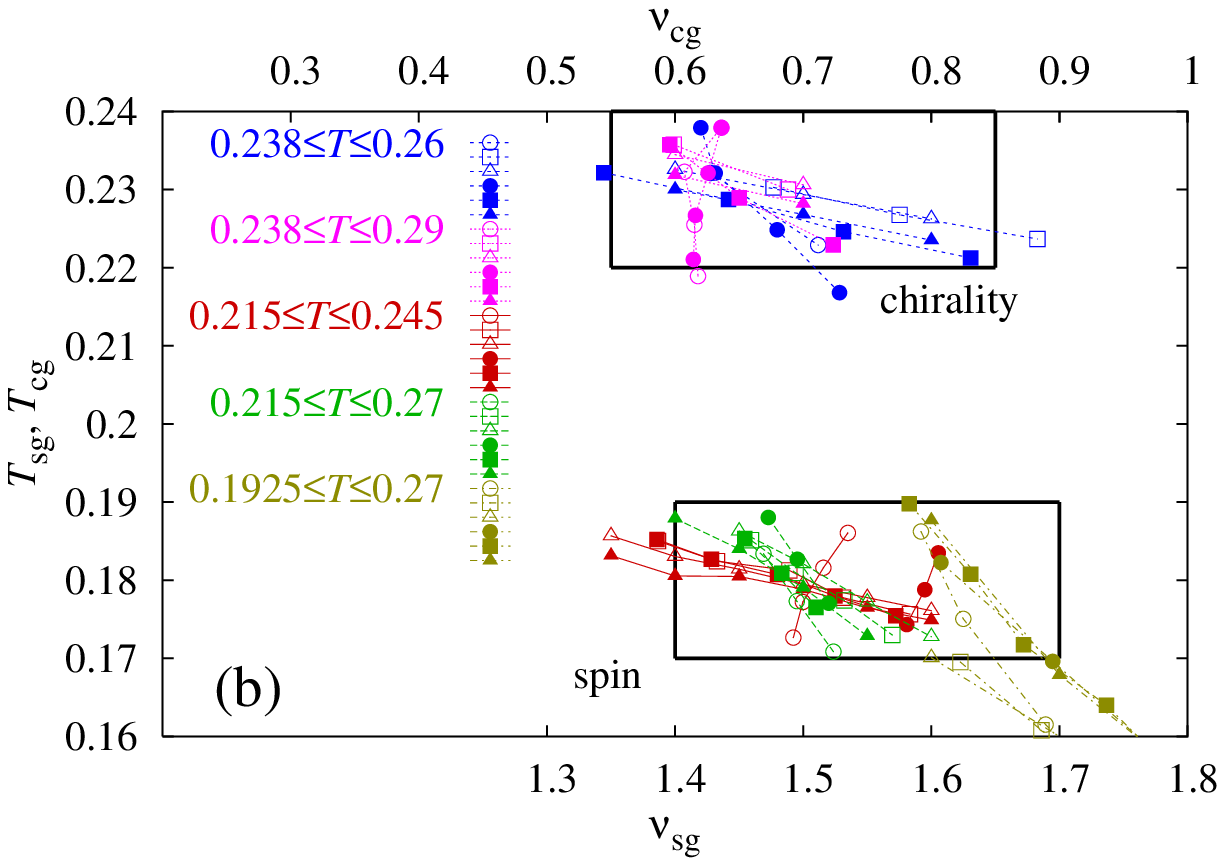}}
  \end{center}
  \caption{
(Color online)
Estimations of finite-time-scaling parameters in the Heisenberg model.
(a) $T_\mathrm{sg/cg}$ versus $2-\eta_\mathrm{sg/cg}$.
(b) $T_\mathrm{sg/cg}$ versus $\nu_\mathrm{sg/cg}$.
Spin-glass(chiral-glass) results are plotted against the lower(upper) axis.
The circle symbols are results of the scaling procedure-1, 
the squares are of the procedure-2, and
the triangles are of the procedure-3.
Unfilled symbols depict data of 1000 discarding steps, and
filled symbols depict those of 3000 steps.
Color distinguishes the temperature range of data used in the scaling
analysis.
Results of $0.1925\le T \le 0.27$ are obtained by discarding 3000 steps
(unfilled symbols) and 10000 steps (filled symbols).
}
\label{fig:scale-Hsg}
\end{figure}
\begin{figure}
  \begin{center}
  \resizebox{0.45\textwidth}{!}{\includegraphics{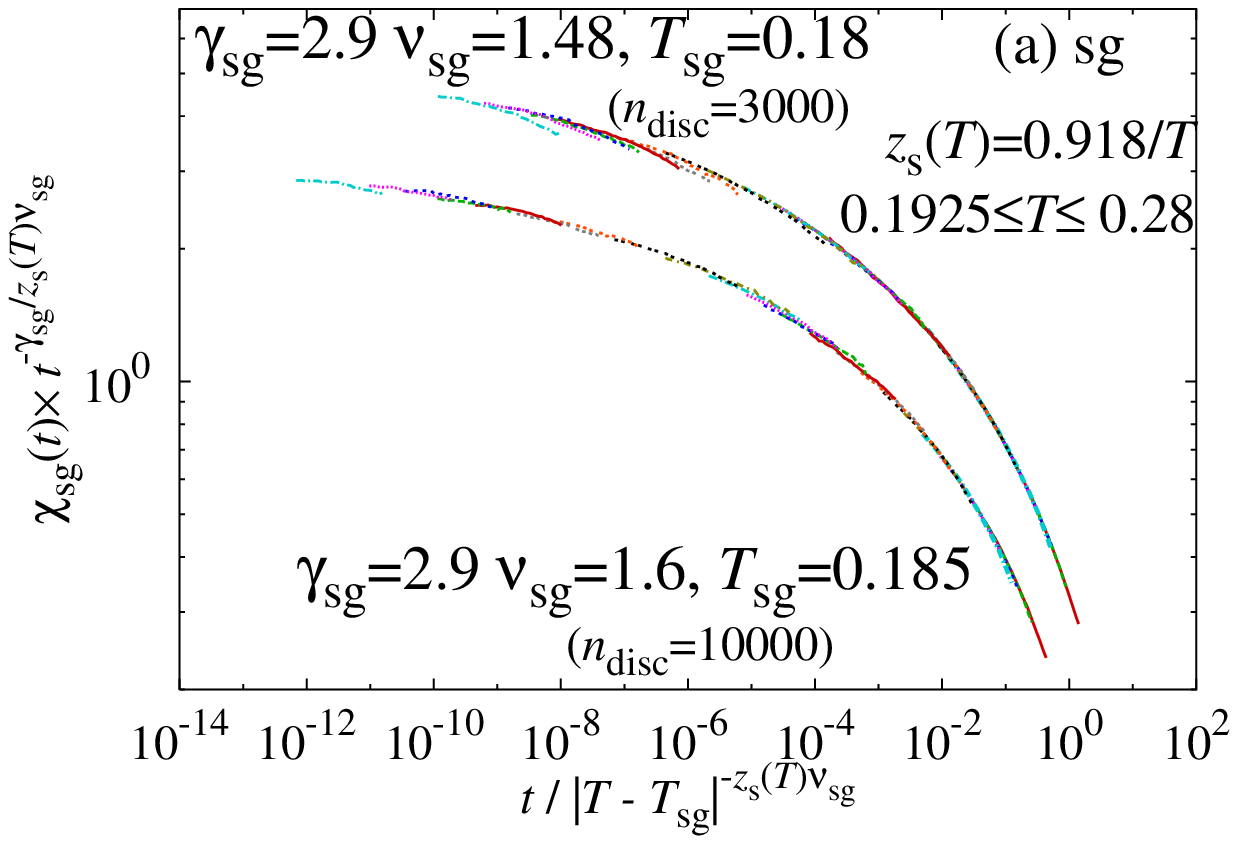}}
  \resizebox{0.45\textwidth}{!}{\includegraphics{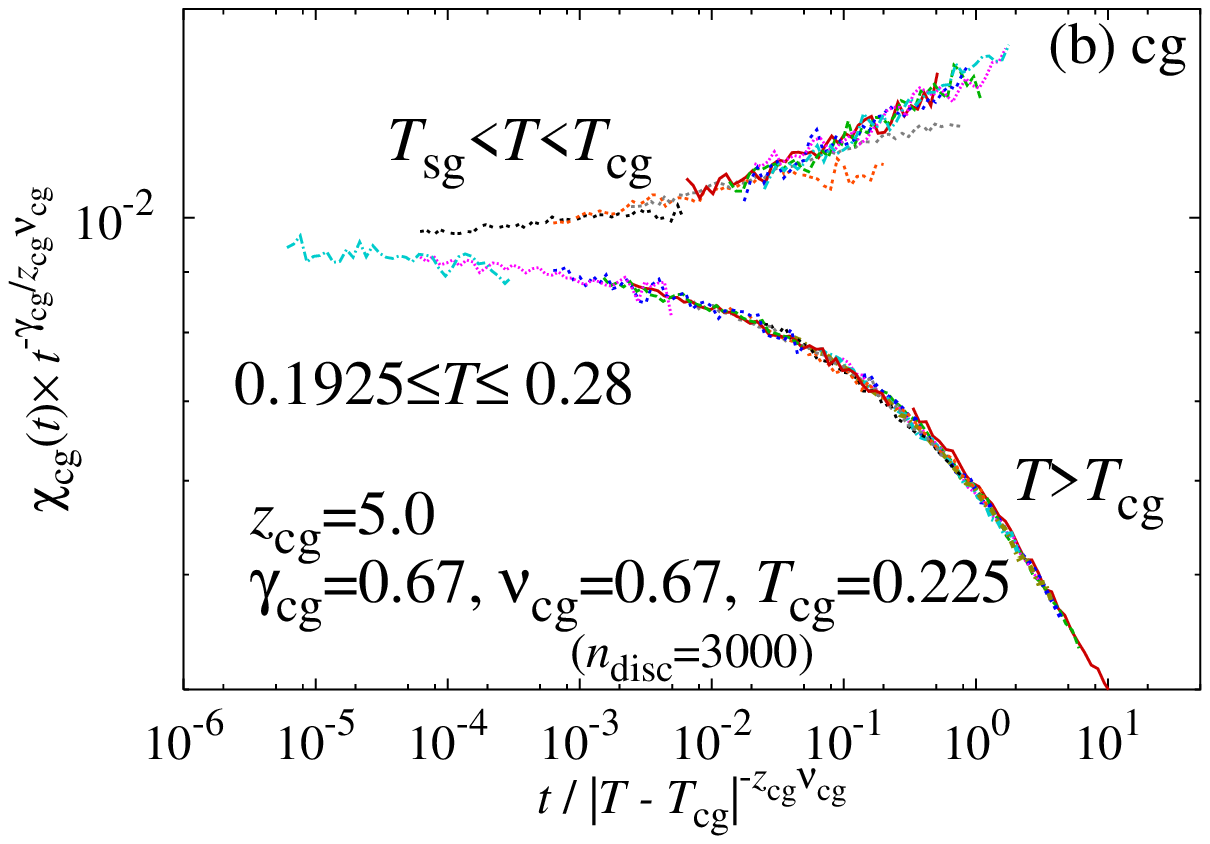}}
  \end{center}
  \caption{
(Color online)
(a)A finite-time-scaling plots of $\chi_\mathrm{sg}$ 
using the estimated parameters.
Two possible chioces of scaling parameters are shown.
As we discard more steps, a value of $\nu_\mathrm{sg}$ increases.
(b) A scaling plot of $\chi_\mathrm{cg}$.
Data of the chiral-glass phase, $T_\mathrm{sg}<T<T_\mathrm{cg}$, are
also scaled using the same parameters.
}
\label{fig:scale-Hsgplot}
\end{figure}

Figures \ref{fig:scale-Hsg} (a) and (b) show estimations of scaling
parameters.
We obtain the transition temperature and the critical exponents, where
the scaling results are robust against the change of procedures.
It is clear from the figures that
the spin-glass transition and the chiral-glass transition occur at 
different temperatures.
Error bars of the critical exponents are rather large compared to those of
transition temperatures.
Particularly, a value of $\nu_\mathrm{sg}$ systematically shifts to a higher 
side when we decrease the lowest temperature of data used in the scaling
analysis from $T=0.215$ to $T=0.1925$.
Then, a value of $\eta_\mathrm{sg}$ is influenced. 
However, a value of $\gamma_\mathrm{sg}$ does not change.
Our estimates for the spin-glass transition are
\begin{eqnarray}
T_\mathrm{sg} &=& 0.18 \pm 0.01\\
\gamma_\mathrm{sg}&=& 2.9 \pm 0.2\\
\nu_\mathrm{sg} &=& 1.55 \pm 0.15\\
\eta_\mathrm{sg} &=& 0.1 \pm 0.1\\
z_\mathrm{sg} &=& 5.1 \pm 0.1,
\end{eqnarray}
and those for the chiral-glass transition are
\begin{eqnarray}
T_\mathrm{cg} &=& 0.23 \pm 0.01\\
\gamma_\mathrm{cg}&=& 0.65 \pm 0.15\\
\nu_\mathrm{cg} &=& 0.70 \pm 0.15\\
\eta_\mathrm{cg} &=& 1.1 \pm 0.1\\
z_\mathrm{cg} &=& 5.0 \pm 0.2.
\end{eqnarray}
In our previous work,\cite{nakamura} 
we assumed that the dynamic exponent does not depend on the temperature.
The obtained results are $ T_\mathrm{sg} = 0.21^{+0.01}_{-0.03}$ and
$ T_\mathrm{cg} = 0.22^{+0.01}_{-0.04}$.
Present estimates roughly correspond to the upper and the lower bound
of the previous estimates.
The error bars shrink, and two transitions are resolved to be different,
which is made possible taking into acount the temperature dependence of
the spin-glass dynamic exponent.
Therefore, the spin-glass results are much influenced.
Results of the chiral-glass transition are consistent with the previous ones,
because an assumption of a constant dynamic exponent is good in this case.
Even though the static exponent, $\nu_\mathrm{sg}$ and $\gamma_\mathrm{sg}$, 
take different values,
$z_\mathrm{sg}$ and $z_\mathrm{cg}$ coincide within the error bars.
They also agree with the dynamic critical exponents of the Ising and the
{\it XY} models.
The spin-glass and the chiral-glass transitions in three dimensions
can be considered to be dynamically universal.

Scaling plots using the obtained results are shown in 
Figs. \ref{fig:scale-Hsgplot}(a) and (b).
For the spin-glass transition, we plot two results with different choices of 
scaling parameters.
They are within the error bars of our estimates.
A value of $\nu_\mathrm{sg}$ tends to increase
when the lowest temperature of the scaling data decreases
and when we increase the discarding steps. 
The transition temperature and a value of $\gamma_\mathrm{sg}$ are rather
robust against these changes.
A scaling plot using a parameter set with $\nu_\mathrm{sg}=1.48$ 
shows some systematic deviations in the long time limit at low temperatures.
Therefore, our best choice in the present analysis is $\gamma_\mathrm{sg}=2.9$,
$\nu_\mathrm{sg}=1.6$, and $T_\mathrm{sg}=0.185$.
A ratio of scaling parameter, $\gamma_\mathrm{sg}/\nu_\mathrm{sg}$, 
agrees with the logarithmic slope of the spin-glass susceptibility of
Fig.~\ref{fig:chisg-Hsg} at $T=T_\mathrm{sg}$.

For the chiral-glass transition, we plot data both above and below the 
chiral-glass transition temperature.
A scaling to a single curve is good above the chiral-glass transition
temperature.
A scaling is also possible for the data in the chiral-glass phase:
$T_\mathrm{sg}<T<T_\mathrm{cg}$.
The scaling function is bending upward, which suggests that a
ratio of critical exponent, $\gamma/\nu$, is increasing as the temperature
decreases.

We also applied the $\beta$-scaling proposed by 
Campbell {\it et. al.}\cite{campbell-huku-taka} and checked our estimates.
As in the Ising model, the critical temperature and the anomalous
exponent $\eta_\mathrm{sg}$ do not change, 
while $\nu_\mathrm{sg}$ and $\gamma_\mathrm{sg}$ increase.
The estimates are 
$T_\mathrm{sg}=0.18$, $\nu_\mathrm{sg}=3.7$, and $\gamma_\mathrm{sg}=6.4$.
Two exponents,
$\nu_\mathrm{sg}$ and $\gamma_\mathrm{sg}$, become particlarly large.
The present numerical data may be insufficient to determine these
critical exponents.
On the other hand,
the spin-glass transition temperature is consistent with our original estimate.
We obtained this value by two scaling analyses and by a temperature dependence
of the chiral-glass dynamic exponent.
We consider that the value is very reliable.

\section{Summary and Discussion}
\label{sec:discussion}

The nonequilibrium dynamics and
the spin-glass transitions in three dimensions are clarified in detail.
It is confirmed that the nonequilibrium dynamic exponent of the spin glasses
continuously depends on the temperature without anomaly at the transition
temperature.
The spin-glass dynamics in the 
high-temperature phase smoothly continues to the low-temperature phase.
The frozen-spin domain grows in the algebraic law controlled by 
$z_\mathrm{s}(T)$ irrespective of the critical temperature.
The growth in the high-temperature phase stops when
the domain size reaches the equilibrium value.
The smooth temperature dependence of the nonequilibrium dynamic exponent 
is considered to be a common characteristic behavior of the spin glasses.
We will need to explain the origin regardless the spin-glass transition,
which is left for the future study.

In the vector spin models, the nonequilibrium dynamic exponent of 
the chiral glass shows a different behavior.
It does not depend on the temperature 
above the ``spin-glass transition temperature".
It remains constant until it changes the behavior at $T=T_\mathrm{sg}$.
When $T<T_\mathrm{sg}$, the chiral-glass dynamic exponent is same with
the spin-glass one.
Therefore, the dynamics of the chiral glass is essentially 
same with the spin-glass dynamics in the low-temperature phase.
They are different only in the high-temperature phase.

Our observation suggests the following scenario.
Dynamics of the spin and the chirality are separated
at temperatures above $T_\mathrm{sg}$.
The dynamic exponent of the spin glass is smaller than the chiral-glass one.
The spin-glass state evolves with time faster than the chiral-glass state.
Therefore, the chiral glass freezes first at a higher temperature.
This is the chiral-glass transition as 
Kawamura\cite{KawamuraXY1,KawamuraXY2,KawamuraXY3,KawamuraH1}
proposed.
Since the spin and the chirality are separated, the spin-glass dynamic exponent
shows no anomaly at $T=T_\mathrm{cg}$.
The situation changes at $T=T_\mathrm{sg}$, where the spin-glass dynamic 
exponent overcomes the chiral-glass dynamic {\it critical} exponent, 
$z_\mathrm{cg}$.
The spin-glass transition occurs at this temperature.
The value, $z_\mathrm{sg}=z_\mathrm{cg}\simeq 5.1$, is found to be common to
all the spin-glass models in three dimensions.
Below the spin-glass transition temperature, 
the spin-glass dynamics is slower than the chiral-glass dynamics
above $T_\mathrm{sg}$, because $z_\mathrm{s}(T)>z_\mathrm{cg}$.
Since the chirality is defined by the spin variables, the chirality obeys the 
spin dynamics once the spin freezes and the spin-glass dynamics becomes 
dominant.
The spin and the chirality is coupled below $T_\mathrm{sg}$, 
and both dynamic exponents show the same temperature dependence.
Therefore, this is the spin-glass phase.

Using the obtained temperature dependences of the dynamic exponent we performed
the finite-time-scaling analysis on the spin- and the chiral-glass 
susceptibility.
The transition temperature and the critical exponents are estimated with
high precisions.
In the Ising model, the results are compared with the previous works
\cite{Bhatt,Ogielski,KawashimaY,palassini,ballesteros,maricampbell}
and the experimental result.\cite{Aruga}
We also checked that a change of a scaling variable may affect a value
of $\nu_\mathrm{sg}$.
In the {\it XY} model, the spin- and the chiral-glass transitions occur
at the almost same temperature.
In the Heisenberg model, the spin- and the chiral-glass transitions are
resolve to be different.
It is the lower and the upper bound of our previous estimate.\cite{nakamura}
Therefore, it becomes clear that the chiral-glass transition without the
spin-glass order occurs in the Heisenberg model, as the 
chirality-mechanism predicted.
\cite{KawamuraXY1,KawamuraXY2,KawamuraXY3,KawamuraH1,HukushimaH,HukushimaH2}
The chiral-glass phase continues for $T_\mathrm{cg}\ge T\ge T_\mathrm{sg}$.

The chirality mechanism insists that the spin-chirality separation is only
observed in the long time and in the large size limit.\cite{HukushimaH2}
A crossover from the spin-chirality coupling regime at short scales to
the decoupling regime at long scales is considered to be important.
Discrepancy regarding whether the simultaneous spin-chirality transition 
occurs or not is explained by the crossover.
It is considered that the true transition will not be observed
unless the simulation goes beyond the crossover length-time scale.
However, we are able to observe the spin-chirality separation near the 
chiral-glass transition temperature within a rather short time 
and a small length scale.
It was made possible by clarifying 
the temperature dependence of the dynamic exponent in the nonequilibrium
relaxation process.
The separation is observed as a different temperature dependence.
If we neglect the effect of the temperature dependence, 
the spin-chirality separation will be smeared out.
Then, we are lead  to the simultaneous transition.

All the dynamic critical exponents at the transition temperatures investigated 
in this paper take almost the same value.
It may be the universal exponent controlling the spin-glass transitions.
Therefore, we consider the dynamical universality about the spin- and the
chiral-glass transitions in three dimensions.

The spin-glass dynamics around the transition temperature 
has exotic behaviors, which are quite different from the regular systems.
A single continuous phase seems to exist in regard with the dynamics.
It should be explained in connection with the picture of the 
low-temperature phase.
It will be a future subject.

\acknowledgments
The author would like to thank
Professor Nobuyasu Ito and Professor Yasumasa Kanada 
for providing him with a fast random number generator RNDTIK.
This work is supported by Grant-in-Aid for Scientific Research from
the Ministry of Education, Culture, Sports, Science and Technology, Japan
 (No. 15540358).

\end{document}